\documentclass[12pt]{iopart}
\usepackage{graphicx}
\usepackage{mathptmx}
\usepackage{flushend}
\usepackage[numbers,sort&compress]{natbib}
\usepackage[colorlinks,citecolor=blue,urlcolor=blue,linkcolor=blue]{hyperref}
\expandafter\let\csname equation*\endcsname\relax
\expandafter\let\csname endequation*\endcsname\relax
\usepackage{amsmath,amssymb}
\usepackage{multirow}
\usepackage{xspace}
\usepackage[caption=false]{subfig}
\usepackage{tikz}
\usepackage{array}

\newcommand\diagg[4]{%
\multicolumn{1}{p{#2}}{\hskip-\tabcolsep
$\vcenter{\begin{tikzpicture}[line width=0.2mm][baseline=1,anchor=south west,inner sep=#1]
\path[use as bounding box] (0,0) rectangle (#2+16\tabcolsep,\baselineskip);
\node[minimum width={#2+16\tabcolsep-\pgflinewidth},
minimum  height=\baselineskip+\extrarowheight-\pgflinewidth] (box) {};
\draw[line cap=round] (0,0.74) -- (9.3, -0.73);
\node[anchor=south west] at (0,-.50) {#3};
\node[anchor=north east] at (9,0.55) {#4};
\end{tikzpicture}}$\hskip-\tabcolsep}}

\begin{document}

\begin{flushright}
ICPP-02
\end{flushright}
%\preprint{ICPP-02}

\title{Constraints on a 2HDM with a singlet scalar and implications in the search for heavy bosons at the LHC}

\author{Stefan von Buddenbrock,$^1$ Alan S. Cornell,$^1$ Elie D. R. Iarilala,$^1$ Mukesh Kumar,$^1$ Bruce Mellado,$^{1,2}$, Xifeng Ruan$^1$ and Esra Mohammed Shrif$^1$}

\address{$^1$School of Physics and Institute for Collider Particle Physics, University of the Witwatersrand, Johannesburg, Wits 2050, South Africa.}
\address{$^2$iThemba LABS, National Research Foundation, PO Box 722, Somerset West 7129, South Africa.}
\ead{\mailto{stef.von.b@cern.ch}, \mailto{alan.stanley.cornell@cern.ch}, \mailto{iarilala@aims.ac.za}, \mailto{mukesh.kumar@cern.ch}, \mailto{bmellado@mail.cern.ch}, \mailto{xifeng.ruan@cern.ch}, \mailto{esra.mohammed.shrif@cern.ch}}
%\vspace{10pt}

\begin{abstract}
We study a two-Higgs doublet model extended with an additional singlet scalar (2HDM+S), and provide a brief introduction to the model and its parameters. Constraints are applied to the parameter space of this model in order to accommodate a number of features in the data that have been interpreted  in Ref.~\cite{vonBuddenbrock:2017gvy} as the result of the $H\rightarrow Sh$ decay produced via gluon-gluon fusion and in association with top quarks. Implications on the phenomenology of the heavy pseudo-scalar ($A$) and charged scalar ($H^+$) are discussed. In particular, the decays $A\rightarrow ZH$ and $H^+\rightarrow W^+H$ become prominent. This leads to final states with multiple leptons and $b$-quarks. The decay $A\rightarrow ZH\rightarrow ZSh$ results in the production of a high transverse momentum $Z$ produced in association with a lepton and two $b$-quarks with little additional jet activity. These predictions are compared to the data with model's benchmark points. With the parameters obtained here the model is able to accommodate the features at the LHC reported in Ref.~\cite{vonBuddenbrock:2017gvy}. Without varying these parameters additional excesses in the $Zb\overline{b}$ and $t\overline{t}$ invariant mass spectra, and the production of $3$ leptons plus two $b$-tagged jets can be explained assuming $m_A\approx 600$\,GeV.
\end{abstract}

\submitto{\jpg}
\noindent{\it Keywords\/}: Higgs boson, singlet scalar, heavy scalar, multiple leptons
%\maketitle

\section{Introduction}
\label{sec:intro}

The discovery of a Higgs boson~\cite{Englert:1964et,Higgs:1964pj,Higgs:1964ia,Guralnik:1964eu} at the Large Hadron Collider (LHC)~\cite{Aad:2012tfa,Chatrchyan:2012xdj} represents a new window of opportunity for the field of particle physics. Following this discovery, the focus has shifted towards the understanding of the couplings of this boson to particles in the Standard Model (SM) and beyond (BSM), and towards the search for new bosons. The Run~2 at the LHC is expected to deliver about 140\,fb$^{-1}$ of usable data at a proton-proton centre of mass energy of 13~TeV. Most of the studies released by the experiments at the LHC have been performed on a quarter of this data set. Furthermore, some important studies have not yet been released with Run~2 data altogether. 

In Refs.~\cite{vonBuddenbrock:2015ema,Kumar:2016vut,vonBuddenbrock:2016rmr} the scalars $H$ and $S$ were introduced via an effective model to explain a number of features in the Run~1 data. These include distortions of the Higgs boson transverse momentum spectrum, accompanied with elevated associated jet activity, elevated rates of leptons in association with $b$-tagged jets used for the search of the associated production of the Higgs boson with top quarks, and results from the search for double Higgs boson and weak boson production. The potential impact on the measurements of some of the signal strengths of the newly discovered Higgs boson has been evaluated in Ref.~\cite{Fang:2017tmh}. The relevance and advantages of electron-proton collisions to search for additional scalar bosons has also been pointed out in Refs.~\cite{Mosomane:2017jcg,Das:2018vuk,DelleRose:2018ndz}.
  
Simple extensions of the Standard Model (SM) are the two-Higgs doublet models (2HDMs)~\cite{Lee:1973iz,Branco:2011iw}, which need an additional Higgs-doublet in the model. As a result of this additional doublet, the scalar spectrum is populated with two CP-even ($h, H$), one CP-odd ($A$) and charged ($H^\pm$) scalar bosons.\footnote{Here we consider $h$ as the lighter Higgs boson as in the SM with $m_h = 125$~GeV.} Note that 2HDMs have been explored in the literature, where various facets related to the theory, phenomenology and constraints on these models using the experimental data from different collider environments have been explored~\cite{Branco:2011iw, Muhlleitner:2017dkd} (and refs. therein). However, as pointed out in Refs.~\cite{Kumar:2016vut,vonBuddenbrock:2016rmr,vonBuddenbrook:2017eqe,vonBuddenbrock:2017jqp}, a 2HDM alone is not able to accommodate the above-mentioned features of the data. As a result, a scalar singlet $S$ is introduced in conjunction with a 2HDM in Ref.~\cite{vonBuddenbrock:2016rmr}, referred to here as the 2HDM+S model. In addition, this type of model may also be able to explore scenarios with dark matter. In Ref.~\cite{vonBuddenbrock:2016rmr} it was discussed that a 2HDM+S model would result in the anomalous production of multiple leptons. This hypothesis has been compared to data~\cite{vonBuddenbrock:2017gvy,Mellado_HDAYS2018}, where large discrepancies between the data and SM Monte Carlos are observed that cannot be resolved with the current understanding of theoretical systematics. The features of the data examined in Ref.~\cite{vonBuddenbrock:2017gvy} have been studied with additional data in Ref.~\cite{vonBuddenbrock:2019ajh}. This study indicates that these features have become more pronounced with more data.

Here we are expand on the phenomenology described in Ref.~\cite{vonBuddenbrock:2017gvy, vonBuddenbrock:2016rmr}. Firstly in this paper we identify the parameter space of the 2HDM+S model that accommodates the features in the data studied in Ref.~\cite{vonBuddenbrock:2017gvy}. Secondly, here we also evaluate the implications of this choice of parameter space for the heavy pseudo-scalar and the charged scalar. We are particularly interested in investigating the CP-odd scalar in the 2HDM+S model. In particular, we study  the production of  $A$ through the gluon-gluon-fusion (ggF) mode, and its decay into $A \to Z H$ channels, where the decay modes $H \to hh, Sh, SS$ are considered. This setup leads to a number of interesting final states with leptons and $b$-tagged jets. 

The resulting kinematics of the decay of $A\rightarrow ZH$, where $m_A > m_Z + m_H$ and $H\rightarrow SS,Sh$, have been studied in Ref.~\cite{vonBuddenbrock:2016rmr}. Assuming that the width of $A$ is much smaller than the experimental resolution obtained with the $llbb$ decay, it was noted that a relatively narrow structure in the invariant mass spectrum of $Zh$ is expected, such that $m_{Zh}<m_A$. 
 
A structure in the $Zh$ invariant mass spectrum has been reported by the ATLAS collaboration~\cite{Aaboud:2017cxo} with Run~2 data. The CMS  collaboration has reported limits with Run~1 and Run~2  data that do not contradict the aforementioned results~\cite{Khachatryan:2015lba,CMS:2018xvc}. The structure has been interpreted in terms of the decay $A\rightarrow Zh$ within a 2HDM, with a cross-section $\sigma(pp\rightarrow Zh)\approx 100-300$\,fb in Ref.~\cite{Ferreira:2017bnx}. Here we interpret the structure in terms of the 2HDM+S model with the spectroscopy discussed in Refs.~\cite{vonBuddenbrock:2015ema,vonBuddenbrock:2016rmr} and Ref.~\cite{vonBuddenbrock:2017gvy}, where $m_H\approx 270$\,GeV and $m_S\approx 150$\,GeV. The structure includes events with more than two $b$-tagged jets, which can be interpreted within a 2HDM as coming from a bottom-quark induced production of the CP-odd boson at intermediate values of $\tan{\beta}$. Here the production of  $Zh$ with additional $b$-tagged jets from $A\rightarrow ZH\rightarrow ZSh,Zhh$ is discussed. 

Note also that the CMS collaboration has reported discrepancies of $2.85\sigma$ in the production of three leptons, where one pair comes from the decay of a $Z$ boson, in association with $b$-tagged jets but with reduced hadronic jet activity~\cite{Sirunyan:2017uzs,Sirunyan:2017leh}. These events are identified in the context of studying $ttZ$ production where the discrepancy emerges with low jet multiplicity. This excess can be explained by the production $A(600)\rightarrow ZH(270)\rightarrow ZS(145)h,Zhh$. In this setup we elaborate on the resulting characteristics corresponding to the production of three leptons, including a $Z$ boson in association with $b$-tagged jets. The potential impact of the signal from a heavy CP-odd boson discussed in Ref.~\cite{Ferreira:2017bnx} is  considered here on the measurement of the signal strength of $Vh$, $(V=ZW)$ production. 

This paper is organised into the following sections: In section~\ref{model} we describe the model, in section~\ref{sec:phenomenology} we describe the tools used and constraints imposed on the parameter scan, reporting on the allowed region of the parameter space where the implications on the branching ratios of the heavy pseudo-scalar and charged scalar are reported. In section~\ref{sec:data} the findings from section~\ref{sec:phenomenology} are compared to the data, and in section~\ref{summ} we summarise and conclude. 

\section{The Model}
\label{model}
Following Ref.~\cite{vonBuddenbrock:2016rmr, Muhlleitner:2016mzt, Ivanov:2017dad}, a 2HDM with an additional real singlet $\Phi_S$ is the baseline for our formalism, where we use the notation used in Ref.~\cite{Muhlleitner:2016mzt}, and call this model the 2HDM+S. As such, the potential is given by:
\begin{align}
V(\Phi_1, \Phi_2, \Phi_S) =&\, m_{11}^2 \left|\Phi_1\right|^2 + m_{22}^2 \left|\Phi_2\right|^2 - m_{12}^2 \left(\Phi_1^\dagger \Phi_2 + {\rm h.c.}\right)
+ \frac{\lambda_1}{2} \left(\Phi_1^\dagger \Phi_1\right)^2 + \frac{\lambda_2}{2} \left(\Phi_2^\dagger \Phi_2\right)^2 \notag \\
&\,  + \lambda_3 \left(\Phi_1^\dagger \Phi_1\right) \left(\Phi_2^\dagger \Phi_2\right)
+ \lambda_4 \left(\Phi_1^\dagger \Phi_2\right) \left(\Phi_2^\dagger \Phi_1\right) + \frac{\lambda_5}{2}\left[\left(\Phi_1^\dagger \Phi_2 \right)^2 + {\rm h.c.}\right] \notag \\
&\, + \frac{1}{2} m_S^2 \Phi_S^2 + \frac{\lambda_6}{8} \Phi_S^4 + \frac{\lambda_7}{2} \left(\Phi_1^\dagger \Phi_1 \right)\Phi_S^2
+ \frac{\lambda_8}{2} \left(\Phi_2^\dagger \Phi_2 \right)\Phi_S^2. \label{pot}
\end{align}
Here the fields $\Phi_1$ and $\Phi_2$ are the $SU(2)_L$ Higgs doublets. The first two lines are the terms from real 2HDM potential, while the last line contains the contribution of the singlet field $\Phi_S$. Generally, models with more than one Higgs doublet have tree-level Flavour Changing Neutral Currents (FCNC). To prevent tree-level FCNCs, we must couple all quarks of a given charge to a single Higgs doublet. This can be accomplished by imposing a $\mathbb{Z}_2$ symmetry, which can be softly broken by the term $m_{12}^2$. Also, the extension of the $\mathbb{Z}_2$ symmetry to the Yukawa sector guarantees the absence of FCNC at tree-level. A trivial generalisation of the usual 2HDM $\mathbb{Z}_2$ 
symmetry requires:
\begin{equation}
 {\Phi}_1 \longrightarrow {\Phi}_1,\quad
 {\Phi}_2 \longrightarrow -{\Phi}_2,\quad
 {\Phi}_S \longrightarrow {\Phi}_S.
\end{equation} 
One can also consider another $\mathbb{Z}_2^{\prime}$ symmetry:
\begin{equation}
 {\Phi}_1 \longrightarrow {\Phi}_1,\quad
 {\Phi}_2 \longrightarrow {\Phi}_2,\quad
 {\Phi}_S \longrightarrow -{\Phi}_S,
\end{equation} 
which is not broken explicitly. For our study, we consider a scenario where the real singlet field $\Phi_S$ acquires a vacuum expectation value ({\it vev}) with $\mathbb{Z}_2$ symmetry.\footnote{In principle $\Phi_S$ could be a complex singlet, and, in this case, the discrete $\mathbb{Z}_2$ symmetry would be promoted to a global $U(1)$ symmetry, where the spontaneous breaking would lead to a massless pseudo-scalar. This might be important and acceptable for phenomenology if it does not couple to the SM particles~\cite{Chen:2013jvg}.} Note that if $\Phi_S$ doesn't acquire a {\it vev}, the $\mathbb{Z}_2^{\prime}$ symmetry then becomes a source of a viable dark matter candidate. 

In this work we set the term $m^2_{12} \neq$ 0 in the 2HDM+S potential, which corresponds to a soft breaking of the $\mathbb{Z}_2$ 
symmetry, and consider the $\lambda_i$ to be real, which corresponds to a model without explicit CP violation.
More discussions can be found in Refs.~\cite{Muhlleitner:2016mzt,Chen:2013jvg,Drozd:2014yla}. 

Assuming the {\it vevs} for the fields $\Phi_1 \to v_1/\sqrt{2}$, $\Phi_2 \to v_2/\sqrt{2}$ and $\Phi_S \to v_S$ are real after electroweak symmetry breaking (EWSB), the minimisation of the potential of the three Higgs fields leads to the three minimisation conditions:
\begin{equation}
 \frac{\partial V}{\partial v_1} =  \frac{\partial V}{\partial v_2} =  \frac{\partial V}{\partial v_S} = 0.
\end{equation}
The first derivative conditions for $\Phi_i$ ($i = 1, 2, S$) are:
\begin{eqnarray}
 \frac{\partial V}{\partial \Phi_1} & = 0 \rightarrow& m^2_{11} = - \frac{1}{2}(v^2_1\lambda_1 + v^2_2\lambda_{345} + v^2_S\lambda_7) + \frac{v_2}{v_1}m^2_{12} \label{e1} , \\
 \frac{\partial V}{\partial \Phi_2} & = 0 \rightarrow& m^2_{22} = - \frac{1}{2}(v^2_2\lambda_2 + v^2_1\lambda_{345} + v^2_S\lambda_8) + \frac{v_2}{v_1}m^2_{12}  \label{e2}, \\
  \frac{\partial V}{\partial \Phi_S}  & = 0 \rightarrow& m^2_{S} = - \frac{1}{2}(v^2_1\lambda_7 + v^2_2\lambda_8 + v^2_S\lambda_6) \label{e3} ,  
\end{eqnarray} 
where $\lambda_{345} \equiv \lambda_3 + \lambda_4 + \lambda_5$. Further, the doublet fields $\Phi_1$, $\Phi_2$ and singlet field $\Phi_S$ can be parameterised as:
\begin{align}
\Phi_1 =
\begin{pmatrix}
\phi_1^\pm \\
\tfrac{1}{\sqrt{2}} \left( v_1 + \rho_1 + i\eta_1  \right)
\end{pmatrix}, \qquad
\Phi_2 =
\begin{pmatrix}
\phi_2^\pm \\
\tfrac{1}{\sqrt{2}} \left( v_2 + \rho_2 + i\eta_2  \right)
\end{pmatrix}, \qquad
\Phi_S = v_S + \rho_S, \label{param}
\end{align}
where $\phi_j^\pm (j = 1, 2 )$ are the charged complex fields, $\rho_i$ are real neutral CP-even fields and $\eta_i$ are the real CP-odd fields. By substituting the parametrisation~(\ref{param}) into the Higgs potential~(\ref{pot}), the mass matrices in the gauge basis can be easily obtained from the second derivatives of the fields. Accordingly, the mass-matrix for the charged ($M^2_{H^\pm}$) and CP-odd ($M^2_{A}$) scalar sector will remain as it is in the 2HDM. Using a $2 \times 2$ rotation matrix, given as:
\begin{align}
\begin{pmatrix}
G^\pm \\
H^\pm
\end{pmatrix}
=
\begin{pmatrix}
\cos\beta & \sin\beta \\
-\sin\beta & \cos\beta
\end{pmatrix}
\begin{pmatrix}
\phi_1^\pm \\
\phi_2^\pm
\end{pmatrix},
\end{align}
where $G^\pm$ are a pair of charged Goldstone bosons and $H^\pm$ are the physical charged scalars; the mass squared charged scalar matrix is the same as in a 2HDM. It can be represented as:
\begin{align}
M^2_{\rm H^\pm} =
\begin{pmatrix}
- \left(2 m_{12}^2 + \left(\lambda_4 + \lambda_5\right) v_1 v_2 \right) \frac{v_2}{2 v_1} & m_{12}^2 + \frac{1}{2} \left(\lambda_4 + \lambda_5\right) v_1 v_2 \\
m_{12}^2 + \frac{1}{2} \left(\lambda_4 + \lambda_5\right) v_1 v_2 & - \left(2 m_{12}^2 + \left(\lambda_4 + \lambda_5\right) v_1 v_2 \right) \frac{v_2}{2 v_1} 
\end{pmatrix}.
\end{align}
Similarly, the CP-odd scalar sector can be diagonalised using the same $2 \times 2$ rotation matrix:
\begin{align}
\begin{pmatrix}
G^0 \\
A
\end{pmatrix}
=
\begin{pmatrix}
\cos\beta & \sin\beta \\
-\sin\beta & \cos\beta
\end{pmatrix}
\begin{pmatrix}
\eta_1 \\
\eta_2
\end{pmatrix},
\end{align}
where $G^0$ is a neutral Goldstone boson, $A$ is the physical pseudo-scalar, and $\eta_{1,2}$ are real CP-odd fields. The mass squared CP-odd scalar matrix is exactly the same as in the 2HDM. It can be formulated as:
\begin{align}
M^2_{A} =
\begin{pmatrix}
- \left( m_{12}^2 + \lambda_5 v_1 v_2 \right) \frac{v_2}{v_1} & m_{12}^2 + \lambda_5 v_1 v_2 \\
m_{12}^2 + \lambda_5 v_1 v_2 & - \left( m_{12}^2 + \lambda_5 v_1 v_2 \right) \frac{v_2}{v_1}
\end{pmatrix},
\end{align}
where diagonalising the CP-odd mass squared matrix will result in the pseudo-scalars physical mass eigenstates $A$.

Since the 2HDM+S CP-even Higgs sector consists of additional Higgs bosons with respect to the 2HDM, due to the addition of the real scalar singlet, the CP-even neutral Higgs mass matrix is enlarged to a $3\times3$ matrix. In the interaction basis, $(\rho_1, \rho_2, \rho_3)$, it can be written as (using eqs.~(\ref{e1})-(\ref{e3})):
\begin{align}
M^2_{\rm CP-even} =
\begin{pmatrix}
\lambda_1 c_\beta^2 v^2 + t_\beta m_{12}^2  & - m_{12}^2 + \lambda_{345} c_\beta s_\beta v^2  &  \lambda_7 c_\beta v v_S \\
- m_{12}^2 + \lambda_{345} c_\beta s_\beta v^2  & \lambda_2 s_\beta^2 v^2 + m_{12}^2 / t_\beta  & \lambda_8 s_\beta v v_S \\
\lambda_7 c_\beta v v_S & \lambda_8 s_\beta v v_S & \lambda_6 v_S^2
\end{pmatrix},
\end{align}
and it can be diagonalised by an orthogonal $3 \times 3$ matrix ${\cal{R}}$, in terms of mixing angles $\alpha_k$ ($k = 1, 2, 3$), and given as:\footnote{The abbreviations used here are $s_{\alpha_k} \equiv \sin\alpha_k$, $c_{\alpha_k} \equiv \cos\alpha_i$ and $t_\beta \equiv \tan\beta$, where $t_\beta$ is defined 
as $t_\beta = v_2/v_1$ and $v^2 = v_1^2 + v_2^2$. By letting $\alpha_{2,3} \to 0$ and $\alpha_1 \to \alpha + \pi/2$ the 2HDM+S approaches the limit of a 
2HDM with an added decoupled singlet, where $\alpha$ diagonalises the $2\times2$ mass matrix of the CP-even sector.} 
\begin{align}
{\cal{R}} =
\begin{pmatrix}
c_{\alpha_1} c_{\alpha_2} & s_{\alpha_1} c_{\alpha_2} & s_{\alpha_2}  \\
- \left(c_{\alpha_1}s_{\alpha_2}s_{\alpha_3} + s_{\alpha_1}c_{\alpha_3}\right) & c_{\alpha_1}c_{\alpha_3} - s_{\alpha_1} s_{\alpha_2}s_{\alpha_3} & c_{\alpha_2}s_{\alpha_3}  \\
- c_{\alpha_1}s_{\alpha_2}c_{\alpha_3} + s_{\alpha_1}s_{\alpha_3} & -\left(c_{\alpha_1}s_{\alpha_3} + s_{\alpha_1} s_{\alpha_2}c_{\alpha_3}\right) & c_{\alpha_2}c_{\alpha_3} 
\end{pmatrix}.
\end{align}
The three physical mass eigenstates $h, S$ and $H$ in terms of the interaction basis $(\rho_1, \rho_2, \rho_3)$, are given as:
\begin{equation}
  \left({\begin{array}{c}
   h \\
   S \\
   H
   \end{array} } \right) = {\cal{R}} \left({\begin{array}{c}
   \rho_1 \\
   \rho_2 \\
   \rho_S
   \end{array} } \right).
\end{equation} 
Here the mixing angles $\alpha_{1,2,3}$ for the CP-even Higgs states can be constrained within 
 $-{\pi}/{2} < \alpha_{1,2,3} < {\pi}/{2}$,
without loss of  generality. And thus the CP-even mass squared matrix $M^2_{\rm CP-even}$ can be diagonalised using the orthogonal matrix ${\cal{R}}$ as: 
\begin{equation}
 {\cal{R}} M^2_{\rm CP-even} {{\cal{R}}}^T = {\rm diag} (m^2_{h}, m^2_{S}, m^2_{H}). \label{m2}
\end{equation}
\begin{table*}[t]
\renewcommand{\arraystretch}{0.9}
\centering
\begin{tabular}{c|c|c|c}
\hline
	& $u$-type   & $d$-type & leptons \\ \hline
type I  & ${{\cal R}_{i2}}/{s_\beta}$  & ${{\cal R}_{i2}}/{s_\beta}$ & ${{\cal R}_{i2}}/{s_\beta}$   \\ 
type II & ${{\cal R}_{i2}}/{s_\beta}$ & ${{\cal R}_{i1}}/{c_\beta}$ & ${{\cal R}_{i1}}/{c_\beta}$ 	\\
lepton-specific	& ${{\cal R}_{i2}}/{s_\beta}$ & ${{\cal R}_{i2}}/{s_\beta}$ & ${{\cal R}_{i1}}/{c_\beta}$ 	\\
flipped & ${{\cal R}_{i2}}/{s_\beta}$ & ${{\cal R}_{i1}}/{c_\beta}$ & ${{\cal R}_{i2}}/{s_\beta}$ 	\\  \hline
\end{tabular}
\caption{\label{yukawa} The coupling coefficient $c\left( H_i f f\right)$ as defined in eq.~(\ref{lyuk}).}
\end{table*}
It is important to note that this model can connect the anomalies and features seen and discussed in section~\ref{sec:intro}, including the few analyses performed in Refs.~\cite{vonBuddenbrock:2015ema, vonBuddenbrock:2016rmr,vonBuddenbrook:2017eqe,Fang:2017tmh,vonBuddenbrock:2017jqp}, by considering the couplings of additional bosons among themselves, as well as with gauge bosons and fermions. For example, consider the coupling of $H_i = (h, S, H)$ with a pseudo-scalar $A$ and the $Z$ boson, where the Feynman rule is given by:
\begin{equation}
\lambda_{\mu}(H_iZA) = \frac{\sqrt{g^2 + g^{'2}}}{2}(p_{H_i} - p_A)_{\mu} \tilde{c}(H_iZA) ,
\end{equation}
where $g^{\prime}$ is the $U(1)_Y$ gauge coupling with 
$\tilde{c}(hZA) = - c_{\alpha_2}s_{\beta - \alpha_1}$,
$\tilde{c}(SZA) = s_{\beta - \alpha_1}s_{\alpha_2}s_{\alpha_3} + c_{\alpha_3}c_{\beta - \alpha_1}$, and 
$\tilde{c}(HZA) = c_{\alpha_3}s_{\beta - \alpha_1}s_{\alpha_2} - s_{\alpha_3}c_{\beta - \alpha_1}$. 
The four-momenta $p_{H_i}$ and $p_A$ of $H$ and $A$ respectively, are taken as incoming. It is to be noted that the tilde over the coupling factor ($\tilde c$) 
denotes that it is not an effective coupling, since due to no SM counterpart it is not normalised to a corresponding SM coupling. 
Similarly, the trilinear Higgs coupling $hSH$ is given by:
\begin{align}
  \lambda_{hSH} =&\, \frac{1}{v} \Bigg(\mu^2 \Big[(2{\cal{R}}_{12} {\cal{R}}_{13} 
   + {\cal{R}}_{32} {\cal{R}}_{33}) c_{\beta} + ({\cal{R}}_{31} {\cal{R}}_{33} 
   - 3{\cal{R}}_{12} {\cal{R}}_{23} {\cal{R}}_{33} - {\cal{R}}_{21} {\cal{R}}_{23})s_{\beta} \nonumber \\
   &\qquad\quad+ 3{\cal{R}}_{12} {\cal{R}}_{22} \left(\frac{{\cal{R}}_{31}}{c_{\beta}} - \frac{{\cal{R}}_{32}}{s_{\beta}}\right) 
   + 3{\cal{R}}_{13} {\cal{R}}_{23} {\cal{R}}_{31} \frac{s^2_{\beta}}{c_{\beta}}\Big] \nonumber \\
   &\,+ \frac{\sum^3_{i=1} m^2_{H_i}}{v_S} \left[{\cal{R}}_{13} {\cal{R}}_{23}{\cal{R}}_{33}v 
   + {\cal{R}}_{12} {\cal{R}}_{22}{\cal{R}}_{32} \frac{v_S}{s_{\beta}} 
   - {\cal{R}}_{11}({\cal{R}}_{22} {\cal{R}}_{32} + {\cal{R}}_{23} {\cal{R}}_{33})\frac{v_S}{c_{\beta}}\right]\Bigg), 
 \end{align} 
where $\mu^2 = m^2_{12}/(s_{\beta}c_{\beta})$ and ${\cal{R}}_{ij}$ are the elements of the orthogonal matrix ${\cal{R}}$.
In this model the Yukawa Lagrangian is given as:
\begin{align}
{\cal L}_{Y} = - \sum_{i=1}^{3} \frac{m_f}{v} c\left( H_i f f\right) {\bar{\psi}}_f \psi_f H_i, \label{lyuk}
\end{align}
where the coupling coefficient $c\left( H_i f f\right)$ are shown in Table~\ref{yukawa} in terms of mixing matrix elements ${\cal R}_{ij}$ and mixing angle $\beta$ and the couplings of $A$ and $H^\pm$ are same as in the 2HDMs. For further details we refer the reader to Ref.~\cite{Muhlleitner:2016mzt}. 

It is also important to note that the parameter space described in Ref.~\cite{Muhlleitner:2016mzt} and the one chosen for this work (see next section~\ref{sec:phenomenology}) are checked with respect to: (a) theoretical constraints, like tree-level perturbative unitarity, the vacuum stability from global minimum conditions of the 2HDM+S potential and conditions which bound the potential from below;  (b) the experimental constraints from $R_b$~\cite{Haber:1999zh,Deschamps:2009rh} and $B\to X_s \gamma$~\cite{Deschamps:2009rh,Mahmoudi:2009zx,Hermann:2012fc,Misiak:2015xwa}; and (c) the compatibility with the oblique parameters $S, T$ and $U$. 

\begin{figure}[tbp]
\centering 
\subfloat[]{\includegraphics[height=6.5cm]{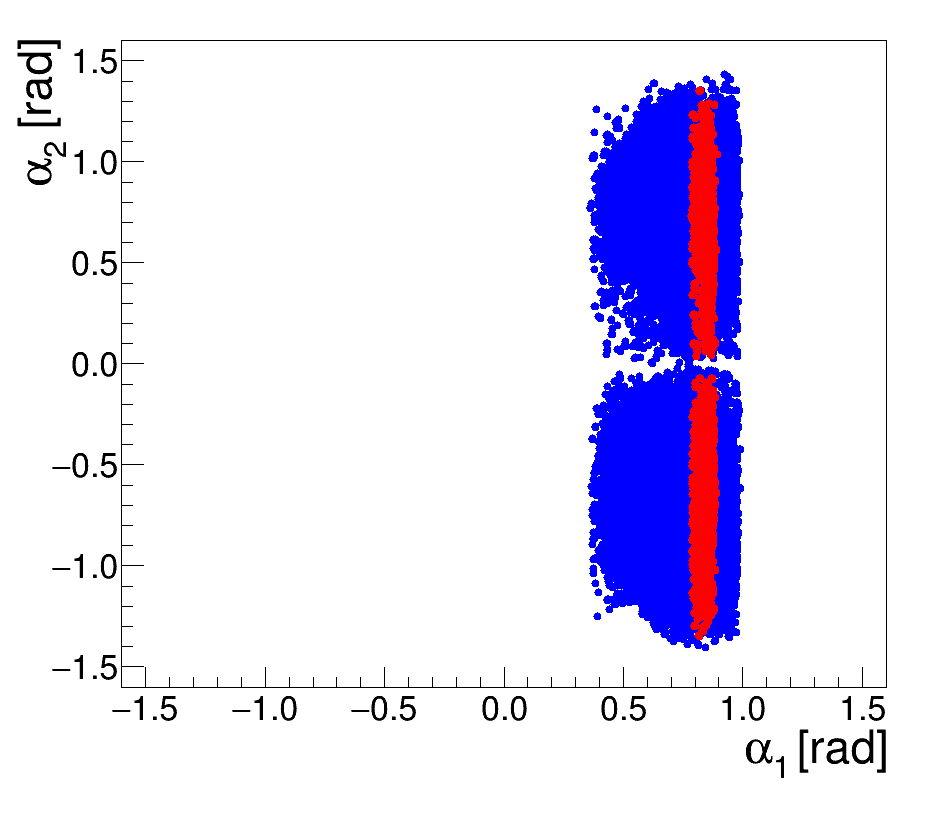}}
\subfloat[]{\includegraphics[height=6.5cm]{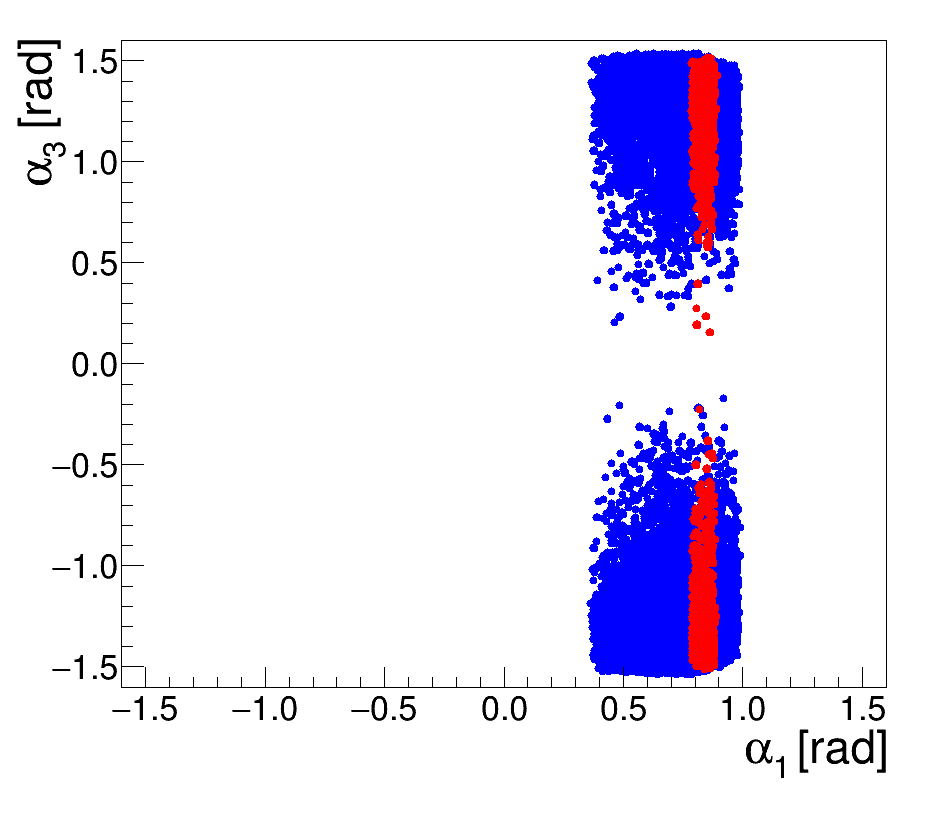} }\\
\subfloat[]{\includegraphics[height=6.5cm]{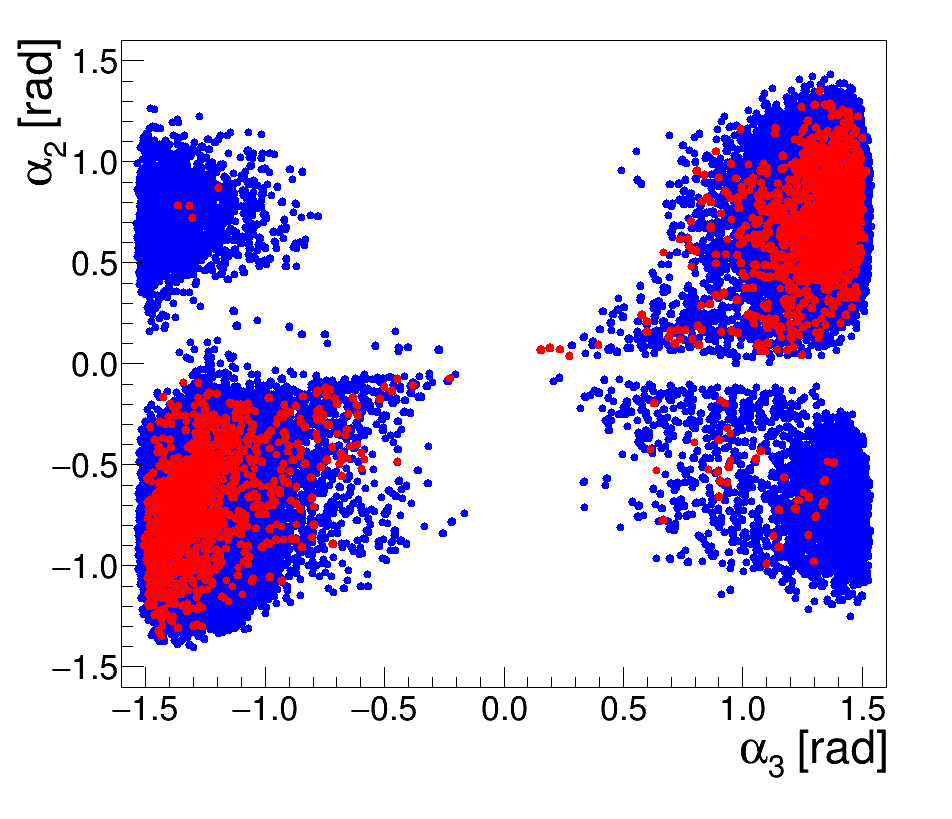}}
\caption{\label{fig:a123} Allowed values of $\alpha_1, \alpha_2$ and $\alpha_3$ for the benchmark considered here, where the values in $red$ $(blue)$ are by considering the BRs of the lightest CP-even scalar consistent within 10\% (20\%) of the prediction for the SM Higgs boson.}
\end{figure} 

\begin{figure}[tbp]
\centering 
\subfloat[]{\includegraphics[height=6.5cm]{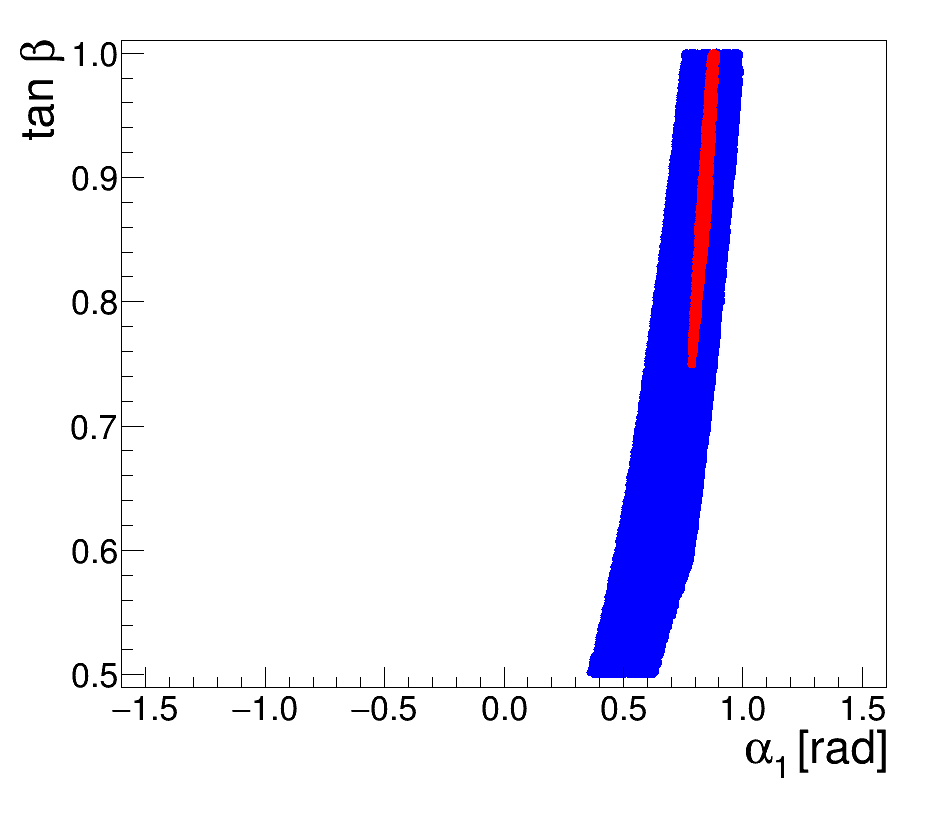}}
\subfloat[]{\includegraphics[height=6.5cm]{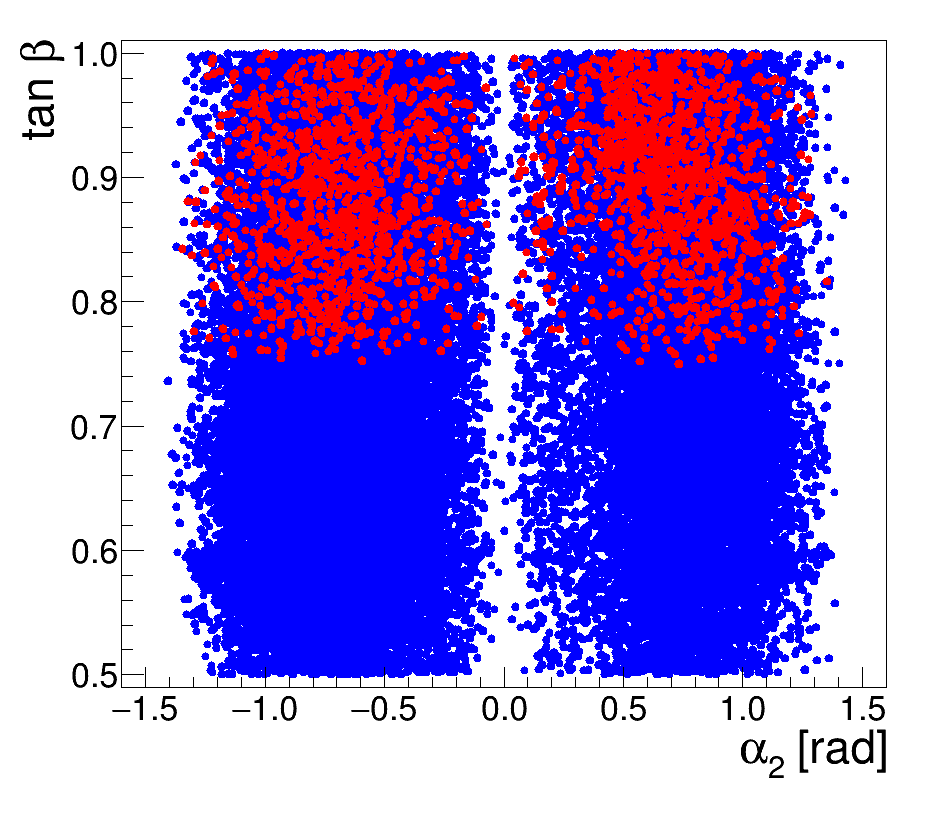}} \\
\subfloat[]{\includegraphics[height=6.5cm]{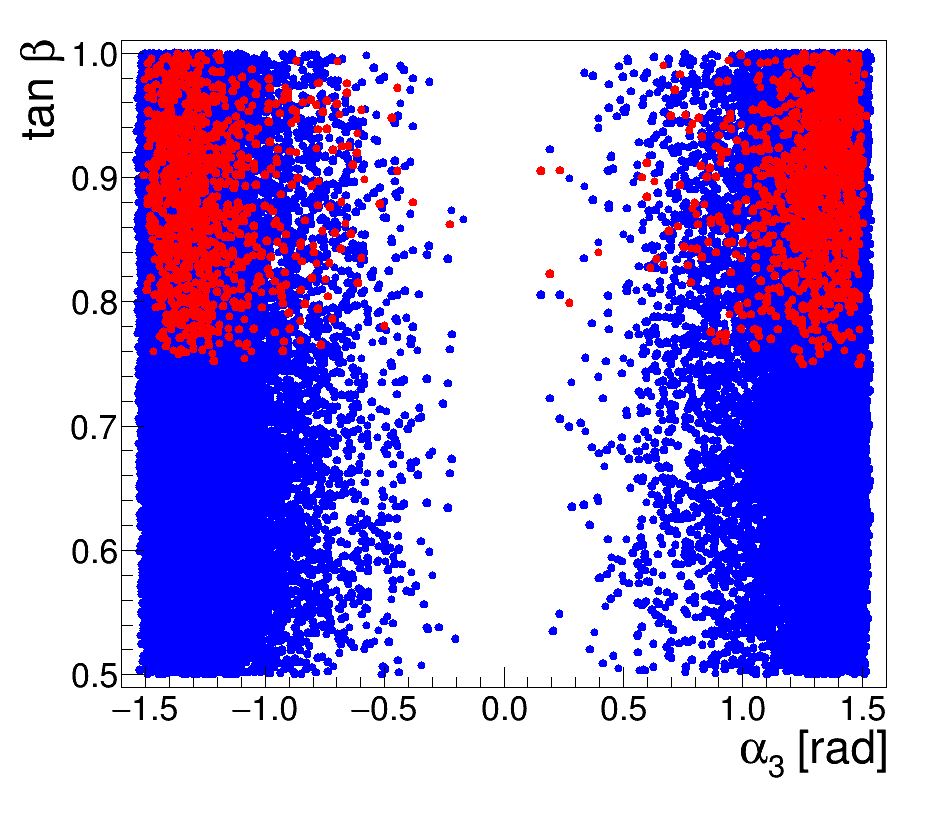}}
\caption{\label{fig:at123} Allowed values of $\alpha_1, \alpha_2$ and $\alpha_3$ against $\tan\beta$ for the benchmark considered here, where the values in $red$ $(blue)$ are by considering the BRs of the lightest CP-even scalar consistent within 10\% (20\%) of the prediction for the SM Higgs boson.}
\end{figure} 

\section{Tools and phenomenology}
\label{sec:phenomenology}
 
In the 2HDM+S model described in section~\ref{model}, the input parameters are following:
\begin{align}
\alpha_1, \alpha_2, \alpha_3, \tan\beta, v, v_S, m_{H_{1,2,3}}, m_A, m_{H^{\pm}}, m_{12}^2.
\end{align}
For our phenomenological analysis we use type-II 2HDM+S throughout and fix the masses of CP-even and CP-odd scalars in the theory as a benchmark point. From here we denote the lightest CP-even scalar as around the SM Higgs boson with mass $m_h=125$\,GeV, the second one as $S$ with mass $m_S=140$\,GeV, and the heaviest one as $H$ with mass $m_H=270$\,GeV. The CP-odd neutral scalar mass $m_A=600$\,GeV and the charged scalar is taken with $m_{H^\pm}=600$\,GeV. These mass values of the scalars are based on the previous studies considered by the authors of Ref.~\cite{vonBuddenbrock:2016rmr}. 
Other parameters are varied in different ranges $-\pi/2 \leq \alpha_{1,2,3} \leq \pi/2$, $0.5\leq \tan\beta \leq 1.0$. Further, we also vary the masses of scalars, $v_S$, $m_{12}^2$ for relevant studies and mentioned at appropriate sections. For numerical calculations we use the publicly available code \texttt{N2HDECAY}~\cite{Muhlleitner:2016mzt}. The \texttt{N2HDECAY} code calculates all 2HDM+S scalar boson decay widths and branching ratios (BRs), which include higher order QCD corrections and off-shell decays, however, in this code electroweak corrections are neglected. Furthermore, theoretical constraints on this model for the above-defined benchmark points, like perturbative unitarity and vacuum stability, are also checked with the package \texttt{ScannerS}~\cite{Coimbra:2013qq}.\footnote{This package is used to perform the checks described in section~\ref{model} and Ref.~\cite{Muhlleitner:2016mzt} in addition with bounds from the collider searches at Tevatron, LEP and LHC.} In addition to these, the following constraints are applied to get the allowed parameter space:
\begin{itemize}
\item BRs of the lightest boson should be consistent with the SM within 20\%  and 10\%. This is in order to comply with the wealth of constraints coming from 
the SM Higgs boson measurements.
\item The sum of the BRs of the heavy scalar to lighter scalars should be 80\%-90\%~\cite{vonBuddenbrock:2017gvy}.
\item The size of the coupling of $h$ to particles in the SM must be in the range $0.8\pm0.12$ of the prediction of the SM~\cite{Mellado_HDAYS2017}.
\item Based on the estimate of the Yukawa coupling of the heavy scalar to top quarks, $\beta_g^2=1.38\pm0.22$~\cite{vonBuddenbrock:2017gvy}, where $\beta_g$ is a scaling factor with respect to the SM. From here it follows $\tan^2{\beta}=0.72\pm0.12$.
\end{itemize}

In Figure~\ref{fig:a123} we show the allowed values of $\alpha_1, \alpha_2$ and $\alpha_3$ for the benchmark considered here. One can appreciate that the mixing angle $\alpha_1$ seems more constrained than the other mixing angles. The constraints on $\alpha_2$ and $\alpha_3$ are better appreciated in Figure~\ref{fig:a123} (c), where one can see that certain areas in the plane are excluded. Results are shown by imposing the condition that the BRs of the lightest scalar be consistent within 20\%  and 10\%. Going from 20\%  to 10\% has a strong impact on how the mixing angles are constrained. Figure~\ref{fig:at123} displays the correlation between mixing angles and $\tan\beta$, where the results are also shown for the two different constraints on the BRs of the light scalar.  The allowed values of $\tan\beta$ become more constrained while going from 20\% to 10\%. The correlation between $\tan{\beta}$ and $\alpha_1$ is noticeable, while the correlation with $\alpha_2$ and $\alpha_3$ appears small.

\begin{figure}[tbp]
\centering 
\subfloat[]{\includegraphics[height=6.5cm]{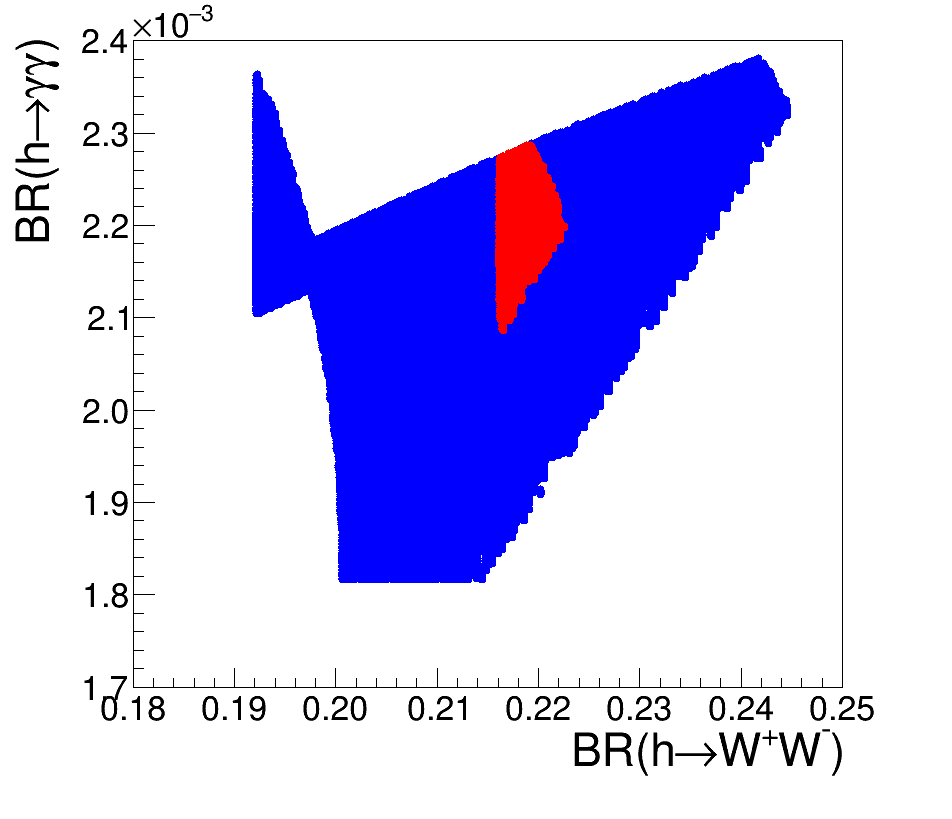}} 
\subfloat[]{\includegraphics[height=6.5cm]{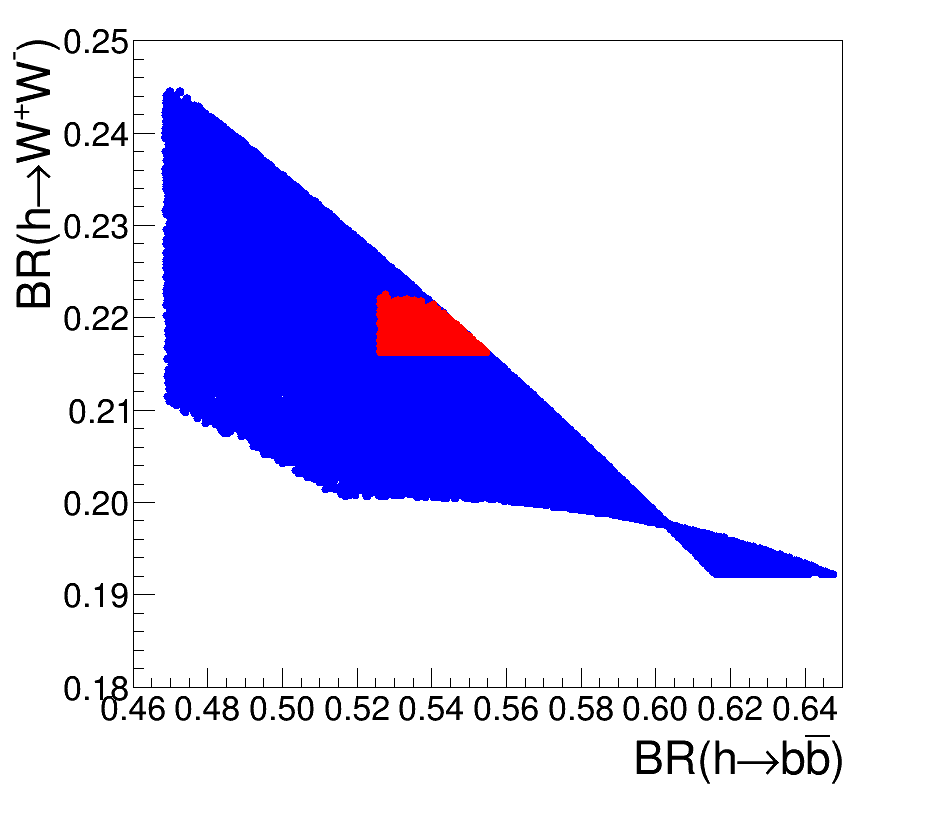}}\\
\subfloat[]{\includegraphics[height=6.5cm]{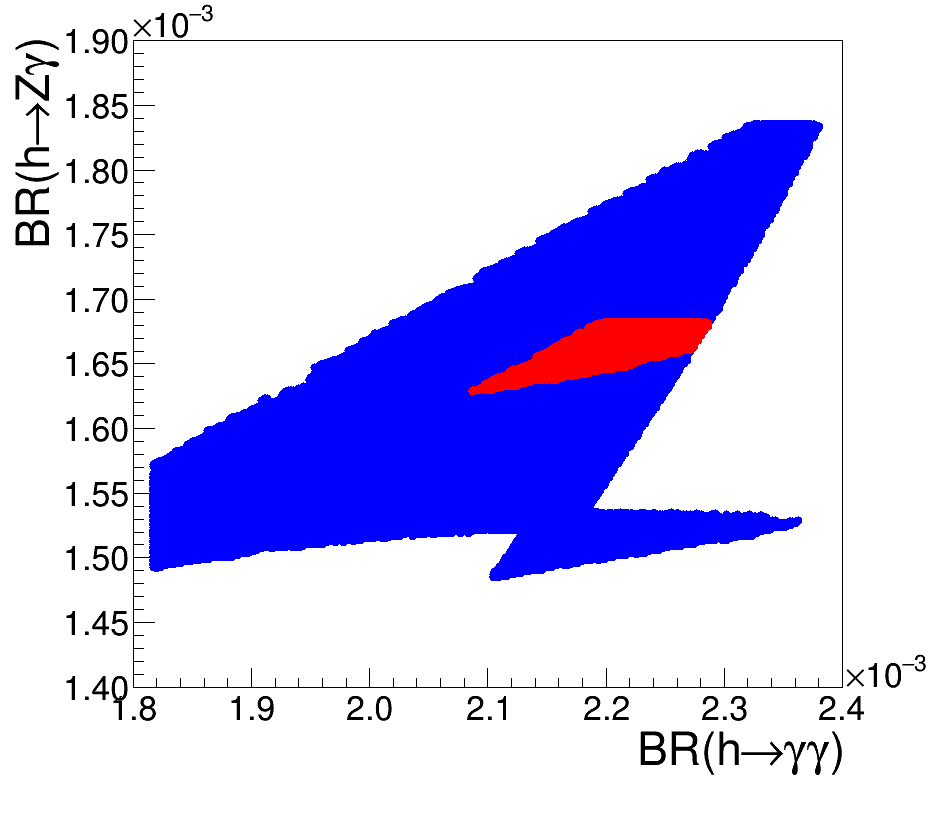}}
\caption{\label{fig:wwh} Correlation plots between the BRs of  
(a) $h\to WW$ \,{\rm vs}\, $h\to \gamma\gamma$,
(b) $h\to b\bar b$ \,{\rm vs}\, $h\to WW$ and 
(c) $h\to \gamma\gamma$ \,{\rm vs}\, $h\to Z\gamma$ where the values in $red$ $(blue)$ are 
by considering the BRs of the lightest $h$ boson consistent within 10\% (20\%).}
\end{figure}

Currently, the sensitivity to the Higgs boson BRs at the LHC is not significantly better than 20\%. This leaves a significant window of opportunity for new physics. Here we investigate the correlations among the relevant BRs that emerge within this 20\% constraint, where ratios of BRs can vary considerably. In Figure~\ref{fig:wwh} we show these correlations for a 20\% as well as for 10\%. One can appreciate that the allowed parameter space corresponding to a maximum 10\% deviation is considerably more constrained than for 20\%. This has to do with the fact that the central values for some of the BRs deviate from the SM, thus strongly restricting the range of deviations from the SM. As the values of the BR departure form the central values of the SM, the blue bands become narrower. While the decay $h\rightarrow Z\gamma$ is not yet observed at the LHC, the correlation between the $h\rightarrow WW$, $h\rightarrow \gamma\gamma$ and $h\rightarrow b\overline{b}$ rates can now be measured by the experiments. The ratio of the BR of $h\rightarrow WW$ to that of $h\rightarrow \gamma\gamma$ will be measured with an accuracy better than 3\%, and an integrated luminosity that is expected to be  accumulated by the High Luminosity LHC. This can be achieved with the application of a full jet veto, where theoretical uncertainties corresponding to the signal production will cancel.

\begin{figure}[tbp]
\centering 
\subfloat[]{\includegraphics[height=6.5cm]{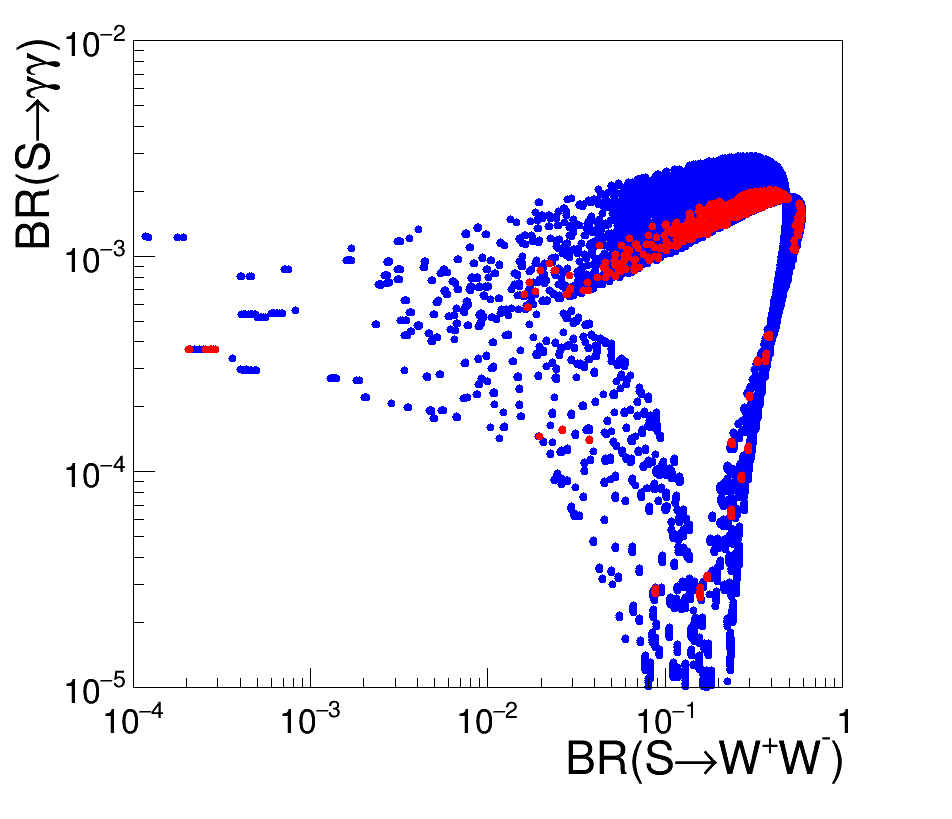}} 
\subfloat[]{\includegraphics[height=6.5cm]{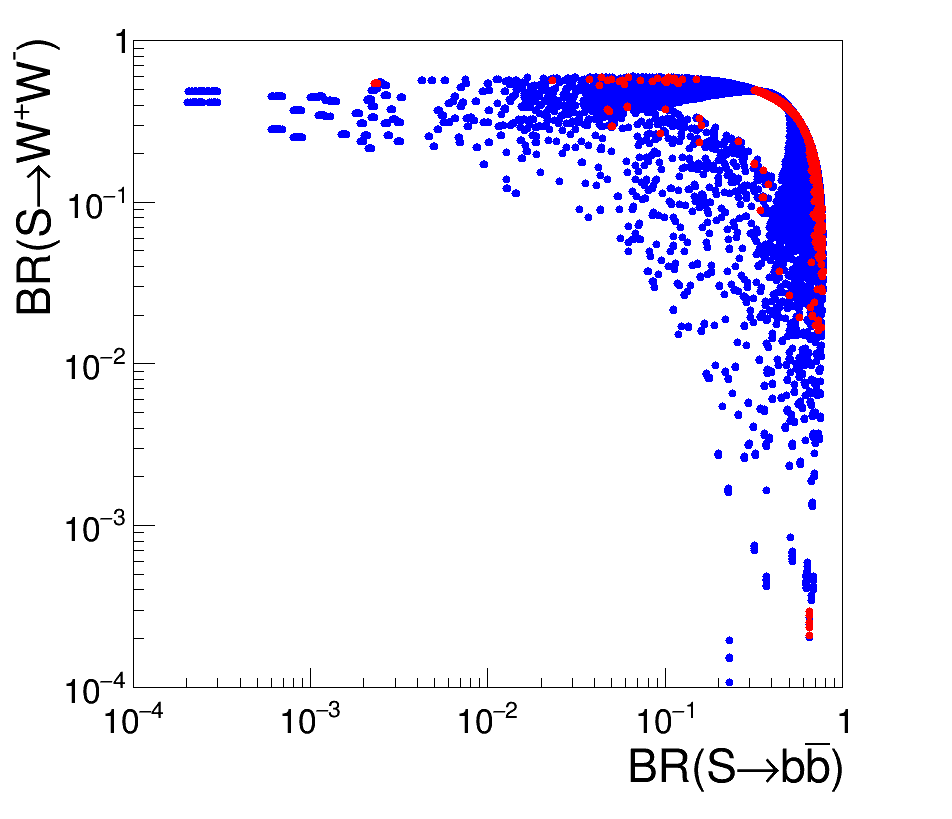}} \\
\subfloat[]{\includegraphics[height=6.5cm]{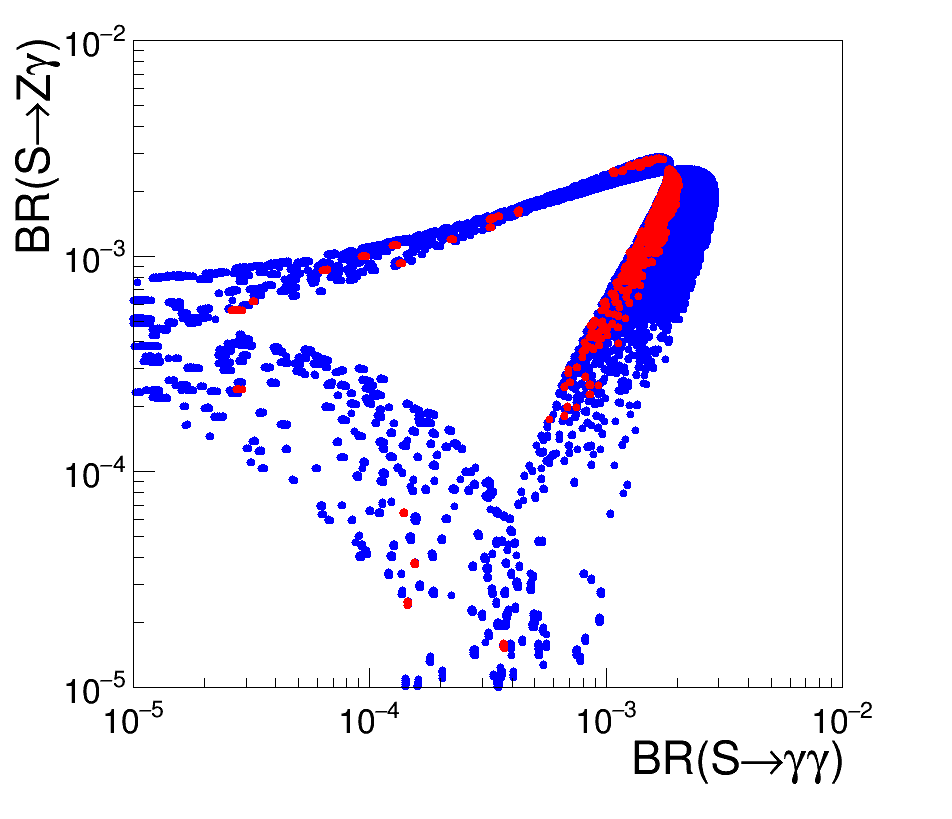}}
\caption{\label{fig:wws} Correlation plots between the BRs of  
(a) $S\to WW \,{\rm vs}\, S\to \gamma\gamma$,
(b) $S\to b\bar b \,{\rm vs}\, S\to WW$ and 
(c) $S\to \gamma\gamma \,{\rm vs}\, S\to Z\gamma$ where the values in $red$ $(blue)$ are 
by considering the BRs of the lightest $h$ boson consistent within 10\% (20\%). Here $m_S=140$~GeV and $m_H=270$~GeV.}
\end{figure}

Figure~\ref{fig:wws} displays the correlations between the BRs of $S$ to SM particles. As opposed to the SM Higgs boson, $S$ is significantly less constrained. Three distinct regimes can be appreciated in Figure~\ref{fig:wws} (b). The first regime corresponds to the dominance of the $S\rightarrow b\overline{b}$ decay, where the second corresponds to the dominance of the $S\rightarrow WW$ decay for $m_S=140$\,GeV. The latter regime would be the preferred one in light of the multi-lepton excesses reported in Ref.~\cite{vonBuddenbrock:2017gvy}. This behaviour is closer to that displayed by a SM Higgs-like boson. In section~\ref{sec:data} $S$ will be assumed to have the same decays as the SM Higgs boson after taking $m_S$ into account. 

The two regimes identified here generate certain correlations in the BR($S\rightarrow \gamma\gamma$) vs BR($S\rightarrow WW$) plane, shown in Figure~\ref{fig:wws} (a), and the BR($S\rightarrow Z\gamma$) vs BR($S\rightarrow \gamma\gamma$) plane, shown in Figure~\ref{fig:wws} (c). The first regime corresponds to the ridges in Figure~\ref{fig:wws} (a) and (c) where the BR($S\rightarrow \gamma\gamma$) is largest. The second and preferred regime corresponds to a situation where BR($S\rightarrow Z\gamma$) $>$ BR($S\rightarrow \gamma\gamma$), with BR($S\rightarrow Z\gamma$)$\approx 10^{-3}$. The latter can be accessible at the High Luminosity LHC~\cite{ATL-PHYS-PUB-2014-006}.

There is a third regime where both BRs in Figure~\ref{fig:wws} (b) display moderate values. In this regime BR($S\to Z\gamma$) and BR($S\to\gamma\gamma$) are small and of order of $10^{-4}$.

\begin{figure}[tbp]
\centering 
\subfloat[]{\includegraphics[height=6.5cm]{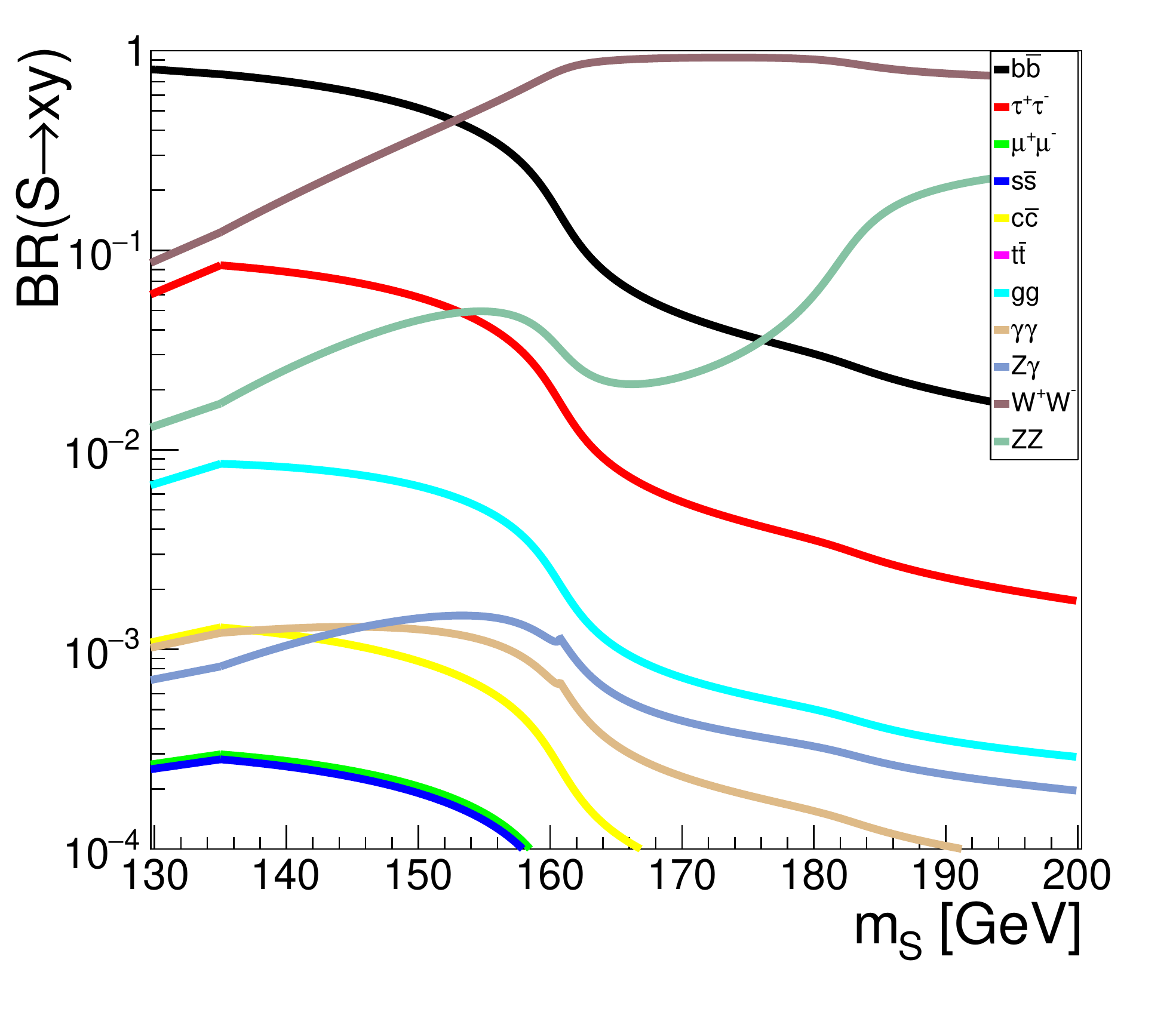}}
\subfloat[]{\includegraphics[height=6.5cm]{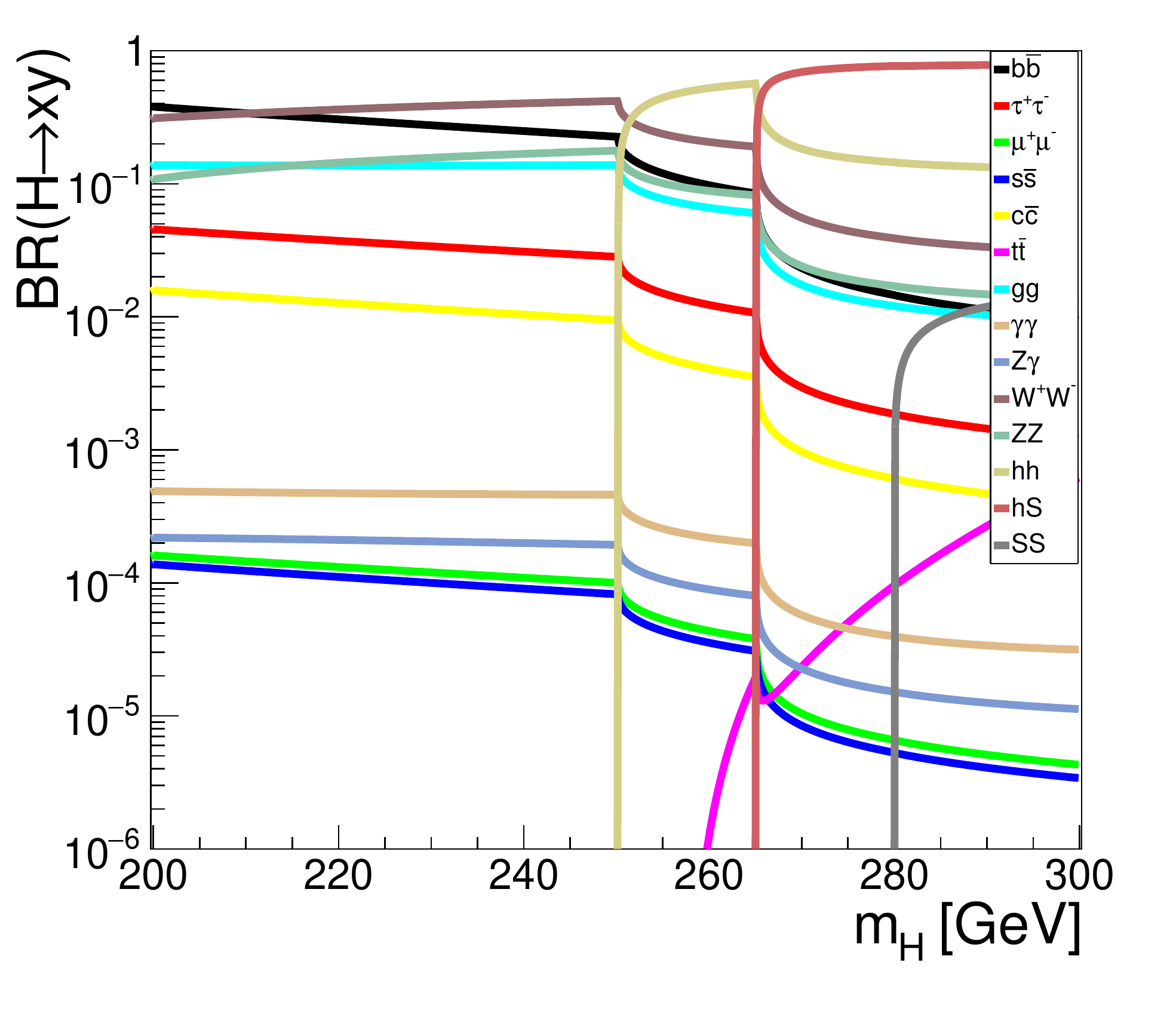}} \\
\subfloat[]{\includegraphics[height=6.5cm]{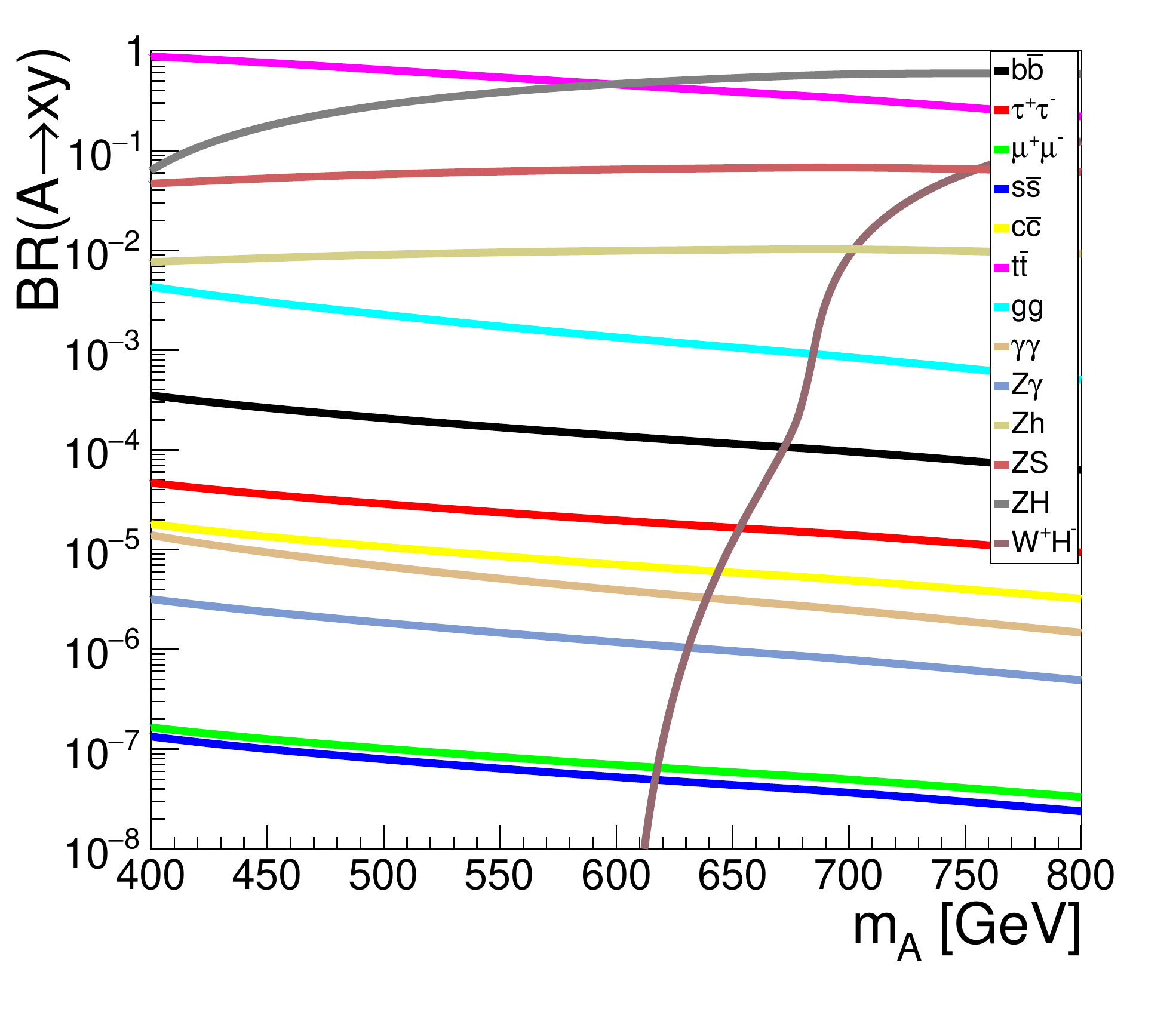}}
\subfloat[]{\includegraphics[height=6.5cm]{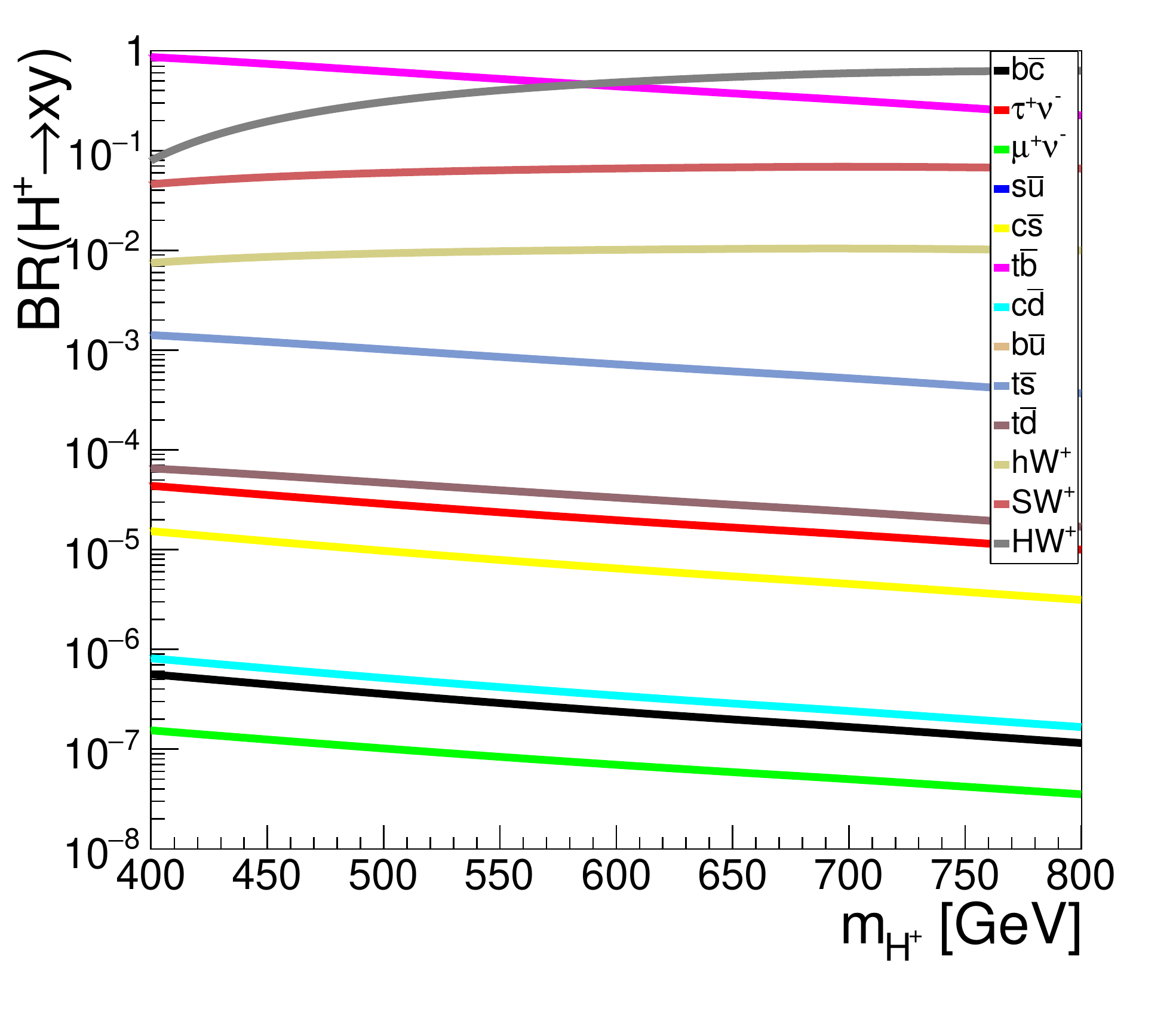}}
\caption{\label{fig:brs} (a) Plot of BR$(S\to x y)$ against $m_S$, (b) BR$(H\to x y)$ against $m_H$. (c) BR$(A\to x y)$ against $m_A$ and (d) BR$(H^\pm \to x y)$ against $m_{H^\pm}$ on the right. In the left, it is shown that the CP-odd scalar $A$ decays predominantly to $ZH$ if its mass is above 600~GeV. On the right, the charged scalar $H^\pm$ decays predominantly to $HW^\pm$ if its mass is greater than 600~GeV.}
\end{figure} 

Furthermore, we investigate the BRs of $S, H, A$ and $H^\pm$ to understand the implications of the constraints detailed above on the decays, and how these can lead to final states that have not been explored before in the context of searches for new bosons. For this purpose a set of allowed parameter values is selected: $\alpha_1 = + 0.885$, $\alpha_2 = - 0.167$, $\alpha_3 = -1.28$, $m_{12}^2 = 3.5$~(TeV$)^2$ and $v_S = 1.5$~TeV. Figure~\ref{fig:brs} shows the BRs of $S, H, A$ and $H^\pm$ in the mass ranges (130 - 200), (200 - 300), (400 - 800) and (400 - 800)~GeV, respectively. This benchmark ensures that the experimental constraints detailed above are described by the model.

Figure~\ref{fig:brs} (a) displays the BRs of the $S$ boson to SM particles, where the mass of $S$ is varied within 
%the range of interest
(130 - 200)~GeV~\cite{vonBuddenbrock:2016rmr}. One can appreciate that the selected set of parameters sit within the regime where the decay to $b\overline{b}$ is dominant for $m_S=140$\,GeV (see the earlier discussions regarding Figure~\ref{fig:wws}). 
For the case when $S$ is treated as a SM Higgs-like boson, the BRs to $W^+W^-$ and $b\overline{b}$ cross just above $m_S=130$\,GeV, where here it happens just over 150\,GeV. On the other hand, the BRs of $Z\gamma$ and $\gamma\gamma$ in the SM Higgs-like boson case crosses at 130\,GeV, but here this happens at 145\,GeV. In the case of $\tau^+\tau^-$ and $ZZ^*$ the corresponding masses are 130\,GeV and just over 150\,GeV, respectively. In this setup the decay to $b\overline{b}$ is dominant up to $m_S=155$\,GeV above which the decay to $W^+W^-$ becomes dominant.

Figure~\ref{fig:brs} (b) displays the BRs of the $H$ boson. It is shown that the CP-even heavy scalar $H$ decays predominantly to $hh$ for $250 < m_H <265$~GeV while above $m_H = 265$~GeV the $H \to Sh$ BR is dominant. These limits on the masses depend on the masses of the other scalars and it is due to the fact that the {\tt N2HDECAY} program does not include  Higgs-to-Higgs off-shell decays in computing the BRs. 
In this setup, and ignoring off-shell decays involving $S$ and $h$ bosons, the dominant decay up to  $m_H \approx 200$\,GeV is $b\overline{b}$, where the decay to $W^+W^-$ overtakes it at $m_H\approx 210$\,GeV. The BRs for rare decays, such as $\gamma\gamma$, $Z\gamma$ and $\mu^+\mu^-$ for $m_H < 2\, m_h$ are of the order of $10^{-4}$. It should be noted that the results from Figures~\ref{fig:brs} (a) and (b) respect the sum rule, which states that the coupling squared of the CP-even scalars to $VV\,(V = W^\pm, Z)$ have to add up to $1$ in terms of the squared SM Higgs coupling to $VV$. 

The constraints from the data implemented here apply to the masses and production rates of the neutral scalar bosons. It is therefore very interesting to evaluate the impact on the BRs of the pseudo-scalar and the charged scalar with the assumption that $m_A,m_{H^{\pm}} > m_H$. Figure~\ref{fig:brs} (c) displays the BRs of the pseudo-scalar. With the parameter choice used here the dominant decay mode in the range $2\, m_t < m_A < 600$\,GeV is $A\rightarrow t\overline{t}$. For $m_A>600$\,GeV the dominant decay is $A\rightarrow ZH$. The latter leads to interesting final states, as discussed in Ref.~\cite{vonBuddenbrock:2016rmr}. In section~\ref{sec:data} we further investigate this decay and, most notably, we scrutinise the rate of production of $ZH$ in association with $b$-tagged jets. In addition, the production of $Z$ in association with a lepton and $b$-tagged jets coming from this decay is compared to the data reported by CMS. The production of this final state with a large enough rate to be produced at the LHC is a feature of the $A\rightarrow ZH$ decay. 
The third most important decay is $A\rightarrow ZS$, which also leads to interesting final states~\cite{vonBuddenbrock:2016rmr}. The decay $A\rightarrow Zh$ is suppressed and sits at the level of 1\%. The production of $Zh$ in association with $S$ and $h$ would come from the decay chain $A\rightarrow ZH\rightarrow ZSh,Zhh$ (see Figure~\ref{fig:brs} (b) and section~\ref{sec:data}). As the pseudo-scalar gets heavier the decay  $A\rightarrow W^\pm H^\mp$ opens up. A distinctive feature of this model is that the decay $A\rightarrow \tau^+\tau^-$ would be suppressed, sitting at the level of $10^{-4}-10^{-5}$, depending on the mass and same follows for the decay $A \to b\bar b$ with a factor of $\sim$10 larger. 

Figure~\ref{fig:brs} (d) shows the BRs of the charged scalar. The decay $H^+\rightarrow t\overline{b}$ is dominant up to $m_{H^+}\approx 600$\,GeV, where the $H^+\rightarrow HW^+$ decay becomes dominant. The latter leads to a tri-boson final state following $H \to h S, h h, W^+W^-, ZZ$ at tree-level. The production of charged Higgs bosons can occur via different modes, $gg$, $q\bar q$ fusion in association with $t$- and $b$-quark at the LHC. Also $H^\pm$ can be produced in association with $t$-quark through the mechanism $gb\rightarrow tH^{-}$. Henceforth, multiple leptons in association with $b$-tagged jets are expected in these production and decay modes of charged Higgs bosons~\cite{vonBuddenbrock:2016rmr}. The third dominant decay is $H^+\rightarrow SW^+$, where as in the case of the $A\rightarrow \tau^+\tau^-$, the $H^+\rightarrow \tau^+\nu$ decay is suppressed. The decay $H^+\rightarrow hW^+$ is suppressed relative to the $H^+\rightarrow HW^+, SW^+$ decays and it stands at about 1\%.

Production cross-sections of different bosons have been checked with the set of parameters used here. The production cross-section of the scalar $S$ is about ten times smaller than that of the SM Higgs boson. The cross-section for the production of $H$ is compatible with that obtained in Ref.~\cite{vonBuddenbrock:2017gvy}. Therefore, the main production mechanism for $S$ would be that of the decay of $H$. 

\section{Comparisons to data}
\label{sec:data}

The primary aim of this section is to confront the data with the benchmark points in the parameter space described in section~\ref{sec:phenomenology}. One of the relevant implications with regards to heavy pseudo-scalars considered here is the dominance of $A\rightarrow t\overline{t}, ZH$ decays, where the decay $A\rightarrow Zh$ appears suppressed. As seen in Figure~\ref{fig:brs}, the branching ratio of the  $A\rightarrow ZH$ decay becomes dominant for $m_A>600$\,GeV.  The decay $A\rightarrow ZH$ leads to interesting final states, as pointed out in Ref.~\cite{vonBuddenbrock:2016rmr}. For the sake of simplicity, here we consider the case where $S$ decays exclusively to SM particles and the decays $H\rightarrow hh,Sh$ are dominant. 

 It is important to reiterate that the scan performed in this section pertains to a significant number of measurements that do not include the $A\rightarrow Zh$ resonance search, which is interpreted here as emerging from $A\rightarrow ZH\rightarrow hh,Sh$. Therefore, the parameters are greatly constrained by data different from the $Zh$ spectrum. The latter constrains the mass of the pseudo-scalar, whereas the branching ratios are constrained with other data sets. 
 
 A distinctive set of final states that emerge from the $A\rightarrow ZH$ decay is the production of relatively high transverse momentum $Z$ bosons in association with a lepton and two $b$-tagged jets. This final state would not appear in $A\rightarrow Zh$ decay in that no additional leptons would be expected in addition to the $Z$ boson and $b$-tagged jets. The $bbA (\to Zh)$  production mechanism could produce this final state. However, the yield would be too small, as discussed below. The CMS experiment has recently reported a discrepancy in the production of $Z$ bosons with an additional lepton, $b$-tagged jets and low additional jet multiplicity~\cite{Sirunyan:2017leh,Sirunyan:2017uzs}. The discrepancy, corresponding to a local significance of $2.85\,\sigma$, appears in the study of the production of $ttZ$ with $Z\rightarrow \ell\ell, \ell=e,\mu$. Here we interpret this discrepancy with the production of $A\rightarrow ZH$, where the parameters of the model are fixed elsewhere, as discussed above.
 
Monte Carlo simulation samples are used to model the background and signal processes for this search. Signal samples are simulated using $\textsc{Pythia8}$~\cite{Sjostrand:2014zea} and then passed  through  $\textsc{Delphes}$~\cite{Ovyn:2009tx} to estimate the detector response. Events were generated for an $A$ boson mass at fixed working points: 500, 550 and 600 GeV. For the interpretation of the $H\to Sh$, where $h$ is the Higgs boson with $m_h$=125 GeV, the masses of the $H$ and $S$ are assumed to be equal to 270 GeV and 145 GeV, respectively. For simplicity, $S$ BRs are taken from that of the SM Higgs boson at the corresponding mass. Jets were clustered using $\textsc{Fast-Jet}$~\cite{Cacciari:2011ma} with the anti-$k_T$ algorithm~\cite{Cacciari:2008gp} using the distance parameter, $R=0.4$. 
 
 Section~\ref{sec:atozh} covers the interpretation of the $A\rightarrow Zh$ search, whereas section~\ref{sec:3lbb} interprets the discrepancy observed in the $Z(\rightarrow \ell\ell) + \ell + 2 b$-tagged jet final state. 

\subsection{$A\rightarrow Zh$ search}
\label{sec:atozh}

The ATLAS and CMS collaborations perform searches for a CP-odd heavy scalar within a 2HDM using various decay channels, where the decay $\ell\ell bb$ plays an important role. A structure in the $Zh$ invariant mass spectrum has been recently reported by the ATLAS collaboration~\cite{Aaboud:2017cxo} with Run 2 data. The CMS collaboration has reported limits with Run 1 data that do not contradict these results~\cite{Khachatryan:2015lba}. 

The authors of Ref.~\cite{Ferreira:2017bnx} have interpreted the structure in terms of a 2HDM where $A\rightarrow Zh$. Here we attempt a different interpretation in light of the spectroscopy discussed in Refs.~\cite{vonBuddenbrock:2015ema,vonBuddenbrock:2016rmr} and Ref.~\cite{vonBuddenbrock:2017gvy} in the context of an extension of the 2HDM, as discussed in section~\ref{model}. In Ref.~\cite{vonBuddenbrock:2016rmr} it was demonstrated that the decay chain $A\rightarrow ZH$, where $H\rightarrow Sh$ generates a structure in the invariant mass spectrum of the $Zh$ system that resembles that of a resonance with moderate width and a mass in the neighbourhood of $m_A-m_S$. The structure reported in Ref.~\cite{Aaboud:2017cxo} peaks around 450\,GeV. Assuming $m_S=145$\,GeV, the mass of the pseudo-scalar considered here is 600\,GeV. The corresponding cross-section of the structure lies in the range between 100\,fb and 300\,fb.

An important feature of the $Zh$ structure is that it also appears in events with additional $b$-tagged jets. Within the context of a 2HDM the structure has been interpreted in terms of the ggF and $bbA$ production mechanisms. In the scenario considered here, the decay entails a three boson final state that also produces a structure in the $Zh$ invariant mass spectrum in association with additional $b$-tagged jets. 

\begin{figure}[tbp]
\centering 
\includegraphics[height=7cm]{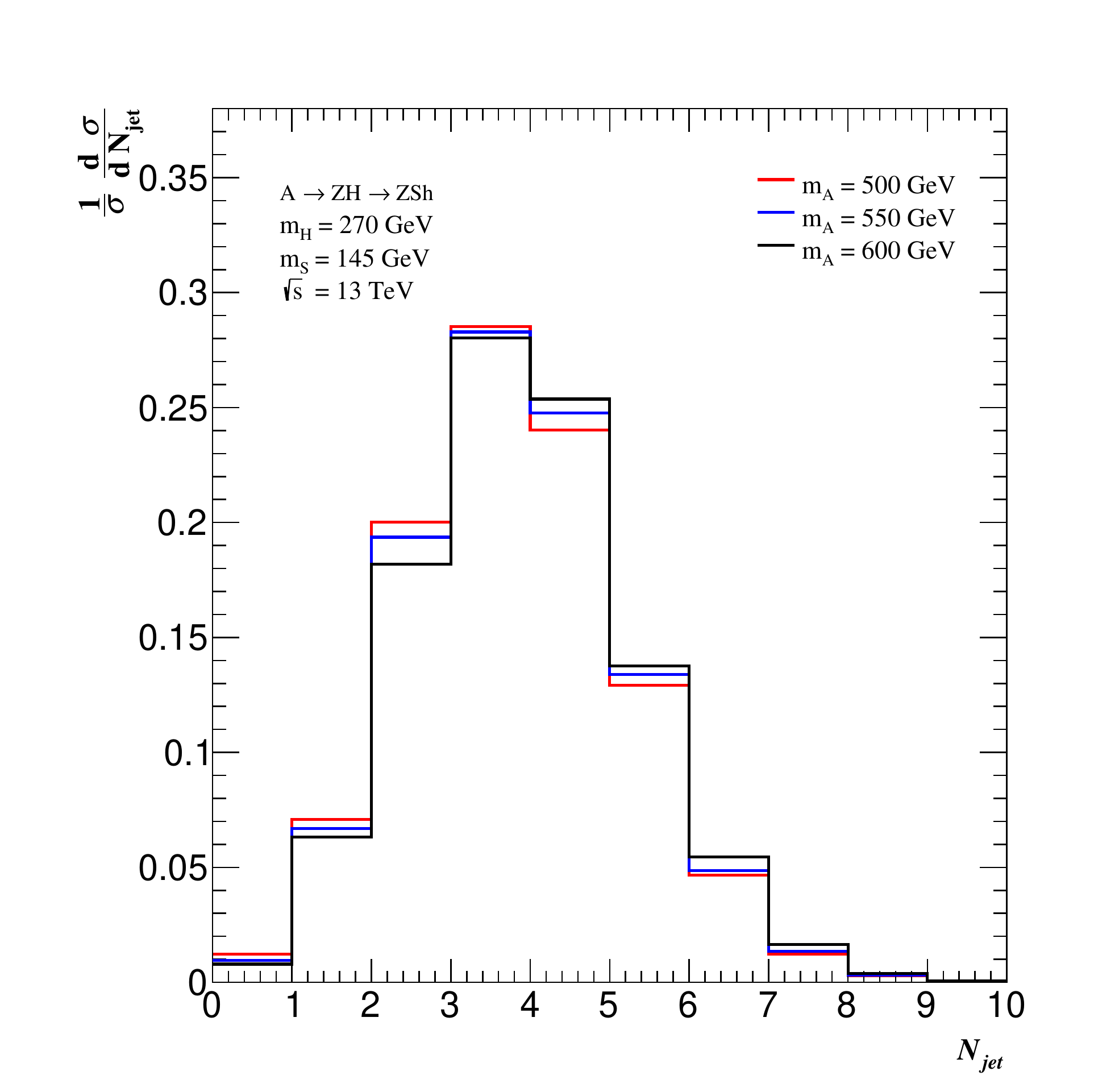}
\includegraphics[height=7cm]{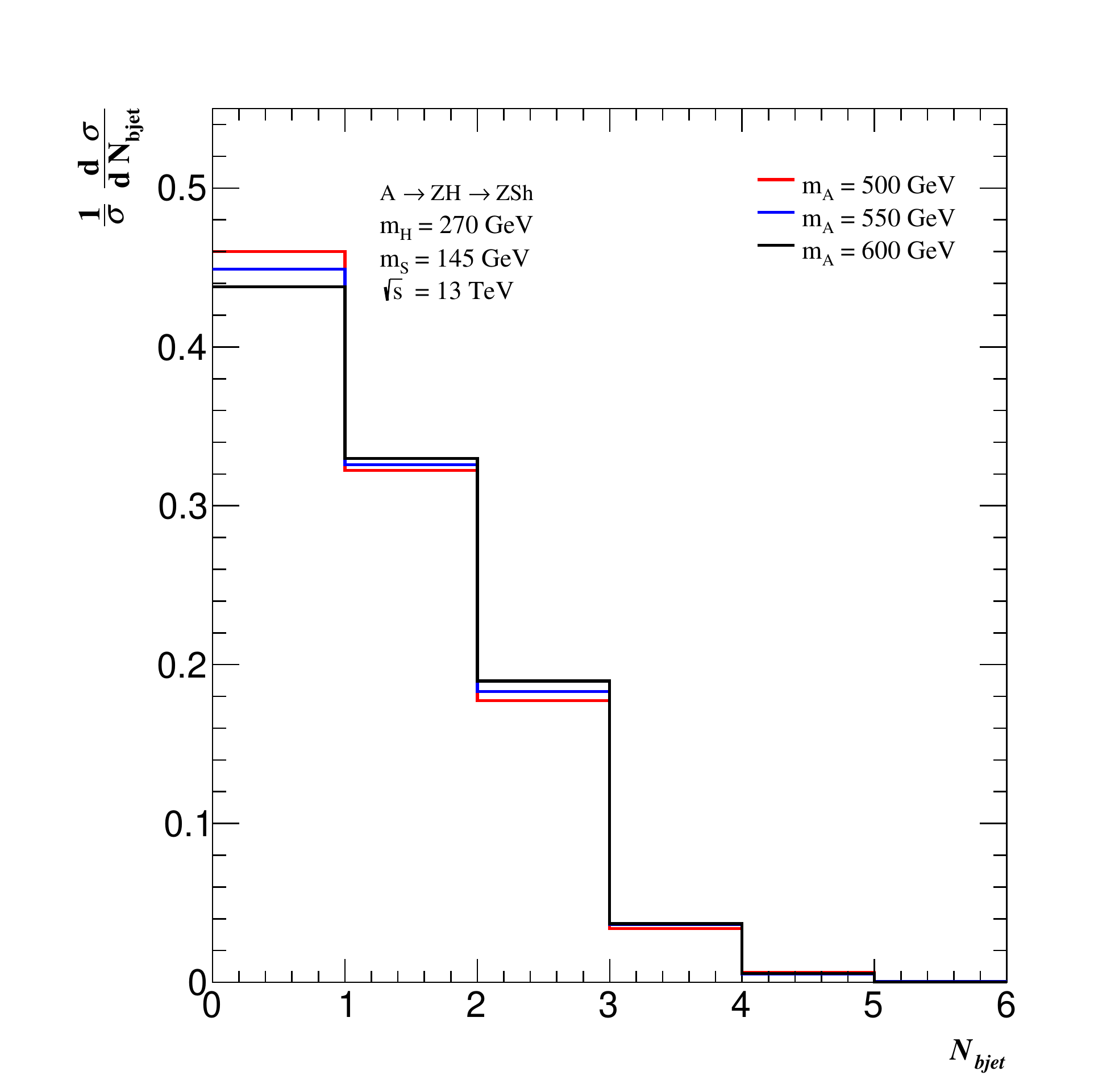} 
\caption{\label{fig:Njet} The jet multiplicity for the decay $A\rightarrow ZH\rightarrow ZSh$ produced via gluon-gluon fusion (see text). The plot on the left corresponds to all jets, whereas the plot on the right shows the multiplicity of $b$-tagged jets.}
\end{figure} 

Figure~\ref{fig:Njet} displays the multiplicity of all jets and $b$-tagged jets for the decay $A\rightarrow ZH\rightarrow ZSh$ produced via ggF. Hadronic jets have $p_T>25$\,GeV and $\left|\eta\right|<2.5$. One can appreciate that the average jet multiplicity overshoots that expected from the direct production of $A\rightarrow Zh$. Because of the significant branching ratio of $S\rightarrow b\overline{b}$, the yield of events with more than two $b$-tagged jets becomes significant so as to mimic the production of $bbA$ in a 2HDM.

In order to evaluate the ability of the 2HDM and the 2HDM+S approaches to describe the data the event selection described in Ref.~\cite{Aaboud:2017cxo} is adopted here as a baseline. This includes the following requirements:
\begin{itemize}
\item At least two $b$-tagged jets are required such that $p_T>20$\,GeV and $\left|\eta\right|<2.5$. The invariant mass of the two leading $b$-tagged jet has to lie in the range $100-145$\,GeV. The transverse momentum of the leading $b$-tagged jet has to be greater than 45\,GeV.
\item It is required to have at least two leptons, the transverse momentum of the leading and sub-leading leptons should be greater than 27\,GeV and 7\,GeV, respectively.
\item The missing transverse momentum is required to be below a threshold (in GeV):
\begin{equation}
E_T^{miss}/\sqrt{H_T} < 1.15 + 8\times 10^{-3}\cdot m_{Zh},
\end{equation}
where $H_T$ is defined as the scalar sum of the transverse momenta of all leptons and hadronic jets. 
\item The transverse momentum of the di-lepton system should conform to the following expression (in GeV):
\begin{equation}
p_{T}^{\ell\ell} > 20 + 9\cdot \sqrt{m_{Zh}-320}
\label{eq:ptll}
\end{equation}
This requirement is applied for $m_{Zh}>320$\,GeV.
\item The invariant mass of the di-lepton system should obey the following expression (in GeV):
\begin{equation}
\text{max}\left[ 40,87-0.030\cdot m_{Zh} \right] < m_{\ell\ell} < 97 + 0.013\cdot m_{Zh}
\end{equation}
\end{itemize}

It is probably relevant to note that this event selection is the result of optimizing for the sensitivity of the search for $A\rightarrow Zh$ in a 2HDM. Here the efficiencies for different production mechanism relative to the ggF production of $A\rightarrow Zh$, in a 2HDM are compared. After the application of the requirements described above the relative rate of additional production of $b$-tagged jets is evaluated. 

\begin{table*}[t]
\renewcommand{\arraystretch}{0.9}
\centering
\begin{tabular}{c|c|c|c|c}
\hline
\textbf{Process} & \textbf{$A(450)\rightarrow Zh$}      & \textbf{$bbA(450)\rightarrow bbZh$} & \textbf{$A(600)\rightarrow Zhh$} & \textbf{$A(600)\rightarrow ZSh$} \\ \hline
\textbf{Efficiency}     & 1    & 1.13 & 1.23 & 0.81  \\ 
\textbf{$f(N_{b}=2)$}	& 0.95 & 0.83 & 0.73 & 0.77	\\
\textbf{$f(N_{b}>2)$}	& 0.05 & 0.17 & 0.27 & 0.23	\\  \hline
\end{tabular}
\caption{\label{tab:1} The efficiency of different production mechanisms of $Zh$ after the application of the event selection described in the text with respect to the ggF production of $A\rightarrow Zh$ in a 2HDM. The second and third rows display the fraction of events with exactly two or more than two $b$-tagged jets after the application of the same event selection. Here $m_H=270$~GeV and $m_S=145$~GeV are used, where $S$ is treated as a SM Higgs-like scalar.}
\end{table*}

Table~\ref{tab:1} shows the efficiency of the different mechanisms for the production of $Zh$ with respect to the ggF production of $A\rightarrow Zh$ in a 2HDM. The efficiency of $A(600)\rightarrow ZH\rightarrow Zhh$ and $A(600)\rightarrow ZH\rightarrow ZSh$ is 23\% larger  and 19\% smaller than that obtained with $A(450)\rightarrow Zh$. The fraction of events with at least one additional $b$-tagged jet varies for different production mechanisms. The fraction is small for $A(450)\rightarrow Zh$, whereas for $bbA(450)\rightarrow bbZh$ the fraction increases to 17\%. The fraction of events with additional $b$-tagged jets increases to 27\% and 23\% for $A(600)\rightarrow ZH\rightarrow Zhh$ and $A(600)\rightarrow ZH\rightarrow ZSh$, respectively. The fraction of events in the $Zh$ mass region between 400\,GeV and 500\,GeV with an additional $b$-tagged jet is about 25\% of the total amount of events in the structure, but the statistical error on this fraction is too large at this moment.

Appendix~\ref{app} provides further details pertaining to the efficiency of $A\rightarrow ZH\rightarrow Zhh$ and $A\rightarrow ZH\rightarrow ZSh$ after the application of cuts used in this section. The final signal efficiency ranges between 0.1\% and 0.35\%, depending on the pseudo-scalar mass and the decay. Appendix~\ref{app} also shows the invariant mass of the $Zh$ system after the application of all cuts. For illustration purposes, for a signal $A(600)\rightarrow ZH\rightarrow ZS(145)h$ and assuming a cross-section of 300\,fb and 36\,fb$^{-1}$ of integrated luminosity, the signal yield would correspond to about 25 events compared to a background of about 100 events. 

\subsection{Three leptons in association with two $b$-tagged jets}
\label{sec:3lbb}

The CMS collaboration has reported a discrepancy in a particular corner of the phase space~\cite{Sirunyan:2017uzs}. This includes the presence of two opposite sign and same flavor charged leptons (electrons or muons) with the invariant mass close to that of the $Z$ boson, an additional charged lepton and at least two $b$-tagged jets. The discrepancy appears in events with exactly two and three hadronic jets. This region of the phase space is weakly populated by the $ttZ$ process, where a large number of hadronic jets is expected. The number of events in the data in excess of the SM prediction corresponds to $28.1\pm 9.5$ events with an integrated luminosity of 35.9\,fb$^{-1}$ at 13\,TeV center of mass energy. 

Following Ref.~\cite{Sirunyan:2017uzs}, the event selection includes the following kinematic requirements:
\begin{itemize}
\item The events are required to have exactly three leptons ($\mu \mu \mu$, $\mu \mu $e, $\mu$ee or eee), where the leading, subleading, and trailing lepton  are required to have $p_{T}$ thresholds above 40, 20, and 10 GeV, respectively.
\item Leptons are required to have $p_T > 10$ GeV and $| \eta | <$ 2.5 (2.4) for electrons (muons).
\item It is required to have a pair of leptons with an opposite charge and same flavor (OSSF) which satisfies $|m_{\ell\ell}- m_{z}| < 10$ GeV.
\item Events containing more than one jet are selected, and then are further split into three categories according to the hadronic jet multiplicity, $N_{j}=2,3,$ and $>3$.
\end{itemize}

Hadronic jets, including $b$-tagged jets, have $p_T>30$\,GeV and $\left|\eta\right|<2.5$. After the application of the event selections described above over the samples we tabulated, the fraction of events for the $A \to Zhh$ and $A\to ZSh$ decay channels are as detailed in Table~\ref{FracEvt}. The quantitative analysis reported here indicates that about 60\% to 65\% of the signal displays low hadronic jet multiplicity with $N_j<3$. 

\begin{table}
\centering
\begin{tabular}{cccccc}
\hline
Process   & $N_j$  & $A$ = 500 GeV& $m_A$ = 550 GeV &$m_A$ = 600 GeV\\
\hline
                                  & =2 &0.26&0.27&0.25\\
$A \rightarrow ZH \rightarrow Zhh$& =3 &0.39&0.38&0.36\\
                                  &$>$3&0.35&0.35&0.39\\
\hline
                                  & =2 &0.32&0.27&0.23\\
$A \rightarrow ZH \rightarrow ZSh$& =3 &0.33&0.36&0.30\\
                                  &$>$3&0.35&0.37&0.46\\
\hline
\end{tabular}
\caption{ The fraction of events after the application of the event selections described in the text with respect to three lepton final states for both $A\to ZH\to Zhh$ and $A\to ZH\to ZSh$ signals. Here $m_H=270$~GeV and $m_S=145$~GeV are used.}
\label{FracEvt}
\end{table}

The efficiency of the event selection with \texttt{DELPHES} is checked against that reported for the production of $ttZ$ in events with at least two $b$-tagged jets and at more than three jets. Taking into account the results from section~\ref{sec:atozh}, the expected yield of the signal $A(600)\rightarrow ZH(270)\rightarrow ZS(145)h$ with $N_j<3$ is about 3 events. This is to be compared to a background of about 50 events. This estimate appears low despite the relatively large uncertainties that characterise the discrepancy discussed here. That being said, it is very important to note that the simplified ansatz that $S$ behaves as a SM Higgs-like boson plays a very important role in the prediction of the cross-section in the corner of the phase-space described here. The assumption made here impacts directly the branching ratio of $S$ decaying into leptons and the jet multiplicity in the final state. The contribution from the $bbA(450)$ signal in this corner of the phase-space is too little to be considered. 

\subsection{Measurement of $Vh$ production signal strength}
\label{sec:vh}

The ATLAS~\cite{Aaboud:2017xsd,ATLAS:2018nkp} and CMS~\cite{Sirunyan:2017elk,CMS:2018abb} collaborations have independently reported observation of the decay of the decay $h\rightarrow b\overline{b}$. The evidence reported is based on excesses in the data in the search for the SM Higgs boson in association with a $Z$ or a $W$ boson. In doing so the experiments also report the corresponding signal strength of the $Vh$ production mechanism in the SM.

Different event selections are developed depending on the presence of high transverse momentum charged leptons. Here the impact of the different production mechanisms normalised to the size of the structure observed by ATLAS in the $Zh$ spectrum on the measurement of the $Vh$ signal strength in the different final states is discussed here. The potential contamination from the different BSM signal production mechanisms in the phase-space of the measurement of $Vh$ production is evaluated. 

Events with two jets tagged as containing $b$-tagged jets and with either zero, one or two charged leptons (electrons or muons) are selected. The lepton candidates are required to have $p_{T}> 7\,{\rm GeV}$ and $|\eta| <2.47~(2.7)$ for electrons (muons). All events are required to have at least two jets with $p_{T}> 20\,{\rm GeV}$ and $|\eta| <2.5$, and exactly two with $100\,{\rm GeV}\leq  m_{bb} \leq 145\,{\rm GeV}$ must pass the $b$-tagging requirement. The $p_{T}$ of the leading $b$-tagged jet is required to be above $45 \,{\rm GeV}$. Events are assigned to zero-, one- and two-lepton channels depending on the number of charged leptons. In the following, the physics objects and the event selection criteria for each channel are described:

\begin{itemize}
\item  The zero-lepton events are required to have  $E_{T}^{\rm miss}> 150\,{\rm GeV}$. The scalar sum of the transverse momenta of the jets in the event, $H_{T}$, is required to be less than 150\,GeV. This is to remove a marginal region of phase 	space in which the trigger efficiency exhibits a small dependence on the jet multiplicity. Also, the following angular selection is applied:
\begin{itemize}
\item  $\Delta\phi\big(\textbf{b}_{1},\textbf{b}_{2}\big) < 140^{\circ}$,
\item $\Delta \phi \big(\textbf{E}_{T}^{\rm miss},\textbf{bb} \big) > 120^{\circ}$,  
\item min $[\Delta\phi\big(\textbf{E}_{T}^{\rm miss},\textbf{jets}\big)] > 30^{\circ}$,
\end{itemize}
where $\textbf{b}_{1}$  and $\textbf{b}_{2}$ are the two $b$-tagged jets forming the Higgs boson candidates dijet system $\textbf{bb}$, and  $\textbf{E}_{T}^{\rm miss}> 150 {\rm GeV}$ is the missing transverse momentum which is defined as the negative vector sum of the transverse momenta of electrons, muons and jets.

\item In the one-lepton channel events are required to contain exactly one electron with $p_{T}> 27\,{\rm GeV}$  or one muon with $p_{T}> 25\,{\rm GeV}$ . With the electron an additional of  $E_{T}^{\rm miss}> 30\,{\rm GeV}$  is applied. 

\item In the two-lepton channel events are required to have exactly two leptons of the same flavor with leading lepton $p_{T}>$ 27 GeV. In dimuon events, the two muons are considered to have opposite-sign charges. The  invariant mass of the dilepton system must be consistent with the $Z$ boson mass, such that $81\, {\rm GeV} <m_{\ell\ell}< 101\,{\rm GeV}$.

\end{itemize}
Events are then categorised into two categories according to jet multiplicity. In the zero- and one-lepton channels events are considered with three or fewer jets. In the two-lepton channel events are considered with higher jet multiplicities which is three or more jets. Furthermore, selections for the reconstructed vector boson's transverse momentum, $p^{V}_{T}$, are applied. This observable corresponds to ${E}_{T}^{\rm miss}$ in the zero-lepton channel, to the vectorial sum of $\textbf{E}_{T}^{\rm miss}$ and  the charged lepton's transverse momentum in the one- and two-lepton channels. In the zero- and one-lepton channels a single region is defined, with $p^{V}_{T} > 150\,{\rm GeV}$. In the 2-lepton channel two regions are considered, $75\,{\rm GeV}< p^{V}_{T} < 150\,{\rm GeV}$ and $p^{V}_{T} > 150\, {\rm GeV}$. Here yields of the different signals are integrated for $p^{V}_{T} > 75\, {\rm GeV}$.

\begin{table*}[t]
\renewcommand{\arraystretch}{0.9}
\centering
\begin{tabular}{c|c|c|c|c}
\hline
\textbf{Process} & \textbf{$A(450)\rightarrow Zh$} & \textbf{$bbA(450)\rightarrow bbZh$} & \textbf{$A(600)\rightarrow Zhh$} & \textbf{$A(600)\rightarrow ZSh$} \\ \hline
$N_{\ell}$=0 & 1.12 & 1.42 & 0.68 & 0.42  \\
$N_{\ell}$=1 & 0.05 & 0.06 & 0.08 & 0.07  \\
$N_{\ell}$=2 & 0.71 & 0.70 & 0.67 & 0.48  \\  \hline
\end{tabular}
\caption{\label{tab:Vhcontamination}  The potential contamination from the BSM production mechanisms discussed here in the measurement of the signal strength of the $Vh,\, Z=Z,W$ in the SM. Results are presented in terms of the signal yield with respect to the $Vh$ production in the SM for zero, one and two charged lepton final states (see text). Here $m_H=270$~GeV and $m_S=145$~GeV are used.}
\end{table*}

 Table~\ref{tab:Vhcontamination} displays the potential contamination of the BSM signals. Results are shown in terms of the BSM signal yield normalized to the yield of the SM $Vh$ production in the phase-space described here. A cross-section of 200\,fb~\cite{Aaboud:2017cxo} is assumed for all BSM signals reported in Table~\ref{tab:Vhcontamination}. It is very important to note that a number of more sophisticated techniques have been used by the experiments to extract the SM signal. As a result, the potential contamination reported in Table~\ref{tab:Vhcontamination} is an upper bound of the potential contamination. Making a more accurate estimate of the contamination goes beyond the scope of this paper. The results shown in Table~\ref{tab:Vhcontamination} are better suited for a comparative analysis. 
 
 One can appreciate that the contamination on the one-lepton final state is quite small, whereas the contamination on the zero-lepton final state is significant, followed by that of the one-lepton final state. The contamination is maximum for the 2HDM signals, where for the $bbA$ production mechanism the contamination would be largest and comparable to the $Vh$ signal in the SM. By contrast, the 2HDM+S signals have a moderate impact on the SM $Vh$ production where the most important decay, $A\rightarrow ZH\rightarrow ZSh$ would have the smallest effect. 

\section{Summary and Conclusions}
\label{summ}

In Refs.~\cite{vonBuddenbrock:2015ema,vonBuddenbrock:2016rmr} scalars $H$ and $S$ were introduced via an effective model to explain a number of features in the Run 1 data. These scalars were embedded into a 2HDM+S model in Ref.~\cite{vonBuddenbrock:2016rmr}, where it was pointed out that the anomalous production of multiple leptons would be a prominent feature of the model. This hypothesis has been compared to data~\cite{vonBuddenbrock:2017gvy,Mellado_HDAYS2018} where large discrepancies between the data and SM MCs are observed that cannot be resolved with available tools. These discrepancies are interpreted using a simplified model where $m_H=270$\,GeV and $m_S=145$\,GeV.

In this paper we attempt to identify the corners of the parameter space in a 2HDM+S model that satisfy the conclusions arrived at in Ref.~\cite{vonBuddenbrock:2017gvy}. The implications on the decays of the heavy pseudo-scalar and charged scalar are discussed. With the parameter choice used here the dominant decay mode in the range $2\, m_t < m_A < 600$\,GeV is $A\rightarrow t\overline{t}$. For $m_A>600$\,GeV the dominant decay is $A\rightarrow ZH$. The third most important decay is $A\rightarrow ZS$, which also leads to interesting final states. The decay $A\rightarrow Zh$ is suppressed and sits at the level of 1\%. The production of $Zh$ would come from the decay chain $A\rightarrow ZH\rightarrow ZSh,Zhh$. As the pseudo-scalar gets heavier the decay  $A\rightarrow W^\pm H^\mp$ opens up. A distinctive feature of this model is that the decay $A\rightarrow \tau^+\tau^-$ would be suppressed, sitting at the level of $10^{-4}-10^{-5}$ depending on the mass. The decay $H^+\rightarrow t\overline{b}$ is dominant up to $m_{H^+}\approx 600$\,GeV where the $H^+\rightarrow HW^+$ decay becomes dominant. The third dominant decay is $H^+\rightarrow SW^+$,  the $H^+\rightarrow \tau^+\nu$ decay is suppressed. The decay $H^+\rightarrow hW^+$ is suppressed relative to the $H^+\rightarrow HW^+, SW^+$ decays. The conclusions from these studies further reinforce the relevance of multi-lepton final states in the search for new bosons. 

A structure in the $Zh$ invariant mass spectrum has been  reported by the ATLAS collaboration~\cite{Aaboud:2017cxo} with Run~2 data, which appears in association with $b$-tagged jets, in addition to those assigned to the decay of $h$. Here we interpret the structure in terms of the 2HDM+S model. The production of  $Zh$ with additional $b$-tagged jets from $A(600)\rightarrow Z(270)H\rightarrow ZS(145)h,Zhh$ is discussed. The fiducial efficiency of this production mechanism is similar to other production mechanisms, like $A(450)\rightarrow Zh$ and $bbA(450)\rightarrow bbZh$. However, one of the features of the 2HDM+S signal considered here is the production of $Z(\rightarrow\ell\ell)$ in association with a lepton and two $b$-tagged jets. The jet activity would be significantly different from that displayed by the production of $ttZ$. An excess is seen by CMS in this final state in events with low additional jet multiplicity, a regime where the production of $ttZ$ is suppressed. By contrast, the 2HDM signals considered here do not contribute significantly to this final state. The potential existence of a heavy pseudo-scalar would contribute to the production of $Zh$ and contaminate the phase-space where the signal strength of the $Vh$ production is measured by the experiments. The 2HDM signals considered here would bring considerable contamination, where the largest contamination would come from $bbA(450)\rightarrow bbZh$, the preferred option in a 2HDM to explain the structure in the $Zh$ invariant mass spectrum. The preferred signal in a 2HDM+S would yield a moderate impact on the measurement of the $Vh$ signal strength. On the other hand, the mass spectrum of this model is computed at tree-level only though the decays are computed with higher-order corrections. So a non-negligible contribution from one-loop corrections to scalars mass may impact the analyses carried here, for example the observables for $A\to ZH$ might get affected. Hence, a future study may be followed considering these corrections.  

As this paper was being reviewed, CMS and ATLAS have reported excesses in $t\overline{t}$ and $Z b\overline{b}$ final states~\cite{Sirunyan:2019wph,CMS:2019wml,Aaboud:2018eoy} that can be interpreted with $m_A\approx 600$\,GeV and in the range of $\tan{\beta}$ considered here.
In conclusion, the 2HDM+$S$ model with the parameters obtained here is able to accommodate the excesses at the LHC reported in Ref.~\cite{vonBuddenbrock:2017gvy}. Without varying these parameters additional excesses in the $Zh$ spectrum and the production of $3$ leptons plus two $b$-tagged jets can be explained assuming $m_A\approx 600$\,GeV.

\section{Appendix}
\label{app}

This appendix reports additional material that is directly relevant to section~\ref{sec:atozh}. This includes an account of the BSM signal efficiency as a function of the pseudo-scalar mass against the event selection used in section~\ref{sec:atozh}. Tables~\ref{tab:cutflow1} and~\ref{tab:cutflow2} display the BSM signal efficiency after successive event selection requirements for the $A\to ZH\to Zhh$ and $A\to ZH\to ZSh$ decays, respectively, where $m_H=270$~GeV and $m_S=145$~GeV. Results are shown for final states with at least and exactly two $b$-tagged jets, as detailed in section~\ref{sec:atozh}. The signal efficiency increases strongly with the pseudo-scalar mass. This driven to a large extent by the requirement in expression~(\ref{eq:ptll}), which was implemented as a result of an optimization for the 2HDM signal discussed in section~\ref{sec:atozh}. As a result, the signal efficiency for $m_A=600$~GeV is about 80\% larger than that for $m_A=500$~GeV. The signal efficiency for the $A\to ZH\to Zhh$ decay is superior to that of $A\to ZH\to ZSh$ for the choice of masses used here. The final signal efficiency ranges between 0.1\% and 0.35\%, depending on the pseudo-scalar mass and the decay chain.

\begin{table*}[h]
\setlength\extrarowheight{5pt}
\centering
%\vspace{0.5cm}
\resizebox{1.\textwidth}{!}{
\begin{tabular}{lcccccc}
\hline
\diagg{.1em}{0.9cm}{{Cuts}}{{Categories}}&\multicolumn{3}{|c|}{{(I) At least 2 $b$-jets}}                                                                          & \multicolumn{3}{c}{{(II) Exactly 2 $b$-jets}}                                                                         \\ \cline{2-7} 
& \multicolumn{1}{|l}{$m_A = 500$\,GeV} & \multicolumn{1}{l}{$m_A = 550$\,GeV} & \multicolumn{1}{l|}{$m_A = 600$\,GeV} & \multicolumn{1}{l}{$m_A = 500$\,GeV} & \multicolumn{1}{l}{$m_A = 550$\,GeV} & \multicolumn{1}{l}{$m_A = 600$\,GeV} \\ \hline 
$N_{bjet}$ = (I) or (II)  & 0.4454 & 0.4504 & 0.4525& 0.2871 & 0.2893 & 0.2904 \\ 
Two leptons & 0.0127 & 0.0143 & 0.0151 & 0.0097 & 0.0108& 0.0115 \\ 
$100<m_{bb}[GeV]<145$ &0.0052 & 0.0055 & 0.0056  & 0.0041 & 0.0043 & 0.0044\\ 
$P_{T}^{led-bj}>45$ GeV & 0.0049 & 0.0053 & 0.0054 & 0.0038 & 0.0041 & 0.0042 \\ 
$P^{\ell\ell}_{T}>20+9\cdot \sqrt{m_{Zh} - 320}$\,GeV & 0.0023 & 0.0036 & 0.0041  & 0.0017 & 0.0026 & 0.0030  \\ 
\small{$\rm{max}[40,87 - 0.030\cdot m_{Zh}] \leq m_{\ell\ell} [\rm GeV] \leq 97 + 0.013\cdot m_{Zh}$} & 0.0021 & 0.0035 & 0.0036  & 0.0015 & 0.0023 & 0.0027 \\ %\cline{1-1}
$E_{T}^{miss}/\sqrt{H_{T}} < 1.15+08 \cdot m_{Vh}$ [GeV] & 0.0020 & 0.0031 & 0.0035 & 0.0014 & 0.0022 & 0.0025\\ \hline
& \multicolumn{6}{c}{$ A\rightarrow  ZH\rightarrow Zhh$}  \\ \hline
\end{tabular}}
\caption{\label{tab:cutflow1}
The fraction of events after the application of the event selections described in section \ref{sec:atozh} with respect to at least and exactly two $b$-tagged jets final states for the signal sample $A\to ZH\to Zhh$, with $m_H=270$~GeV.}
\end{table*}

\begin{table*}[h]
\setlength\extrarowheight{5pt}
\centering
\vspace{0.5cm}
\resizebox{1.\textwidth}{!}{
\begin{tabular}{lcccccc}
\hline
\diagg{.1em}{0.99cm}{{Cuts}}{{Categories}}&\multicolumn{3}{|c|}{{(I) At least 2 $b$-jets}}                                                                          & \multicolumn{3}{c}{{(II) Exactly 2  $b$-jets}}                                                                         \\ \cline{2-7} 
& \multicolumn{1}{|l}{$m_A = 500$\,GeV} & \multicolumn{1}{l}{$m_A = 550$\,GeV} & \multicolumn{1}{l|}{$m_A = 600$\,GeV} & \multicolumn{1}{l}{$m_A = 500$\,GeV} & \multicolumn{1}{l}{$m_A = 550$\,GeV} & \multicolumn{1}{l}{$m_A = 600$\,GeV} \\ \hline 
$N_{bjet}$ = (I) or (II)  & 0.3524 & 0.3587 & 0.3600& 0.2467 & 0.2493 & 0.2504 \\
Two leptons & 0.0113 & 0.0122 & 0.0130 & 0.0092 & 0.0100 & 0.0106 \\ 
$100<m_{bb} [GeV]<145$ & 0.0045 & 0.0046 & 0.0047& 0.0037 & 0.0038 & 0.0039 \\ 
$P_{T}^{led-bjet}>45$\,GeV & 0.0042 & 0.0043 & 0.0044 & 0.0034 & 0.0036 & 0.0037 \\ 
$P^{\ell\ell}_{T}>20+9\cdot \sqrt{m_{Zh} - 320}$\,GeV  & 0.0016 & 0.0026 & 0.0029  & 0.0012 & 0.0020 & 0.0024 \\ 
\small{$\rm{max}[40,87 - 0.030\cdot m_{Zh}] \leq m_{\ell\ell} [\rm GeV] \leq 97 + 0.013\cdot m_{Zh}$}  & 0.0014 & 0.0022 & 0.0025& 0.0010 & 0.0017 & 0.0020   \\ %\cline{1-1}
$E_{T}^{miss}/\sqrt{H_{T}} < 1.15+08 \cdot m_{Vh}$ [GeV] & 0.0013 & 0.0021 & 0.0023 & 0.0010 & 0.0016 & 0.0018 \\ \hline
& \multicolumn{6}{c}{$ A\rightarrow  ZH\rightarrow ZSh$}  \\ \hline
\end{tabular}}
\caption{\label{tab:cutflow2} 
The fraction of events after the application of the event selections described in section \ref{sec:atozh} with respect to at least and exactly two $b$-tagged jets final states for the signal sample $A\to ZH\to ZSh$. Here $m_H=270$~GeV and $m_S=145$~GeV are used.}
\end{table*}

\begin{figure}
\centering 
\includegraphics[width=18pc,height=17pc]{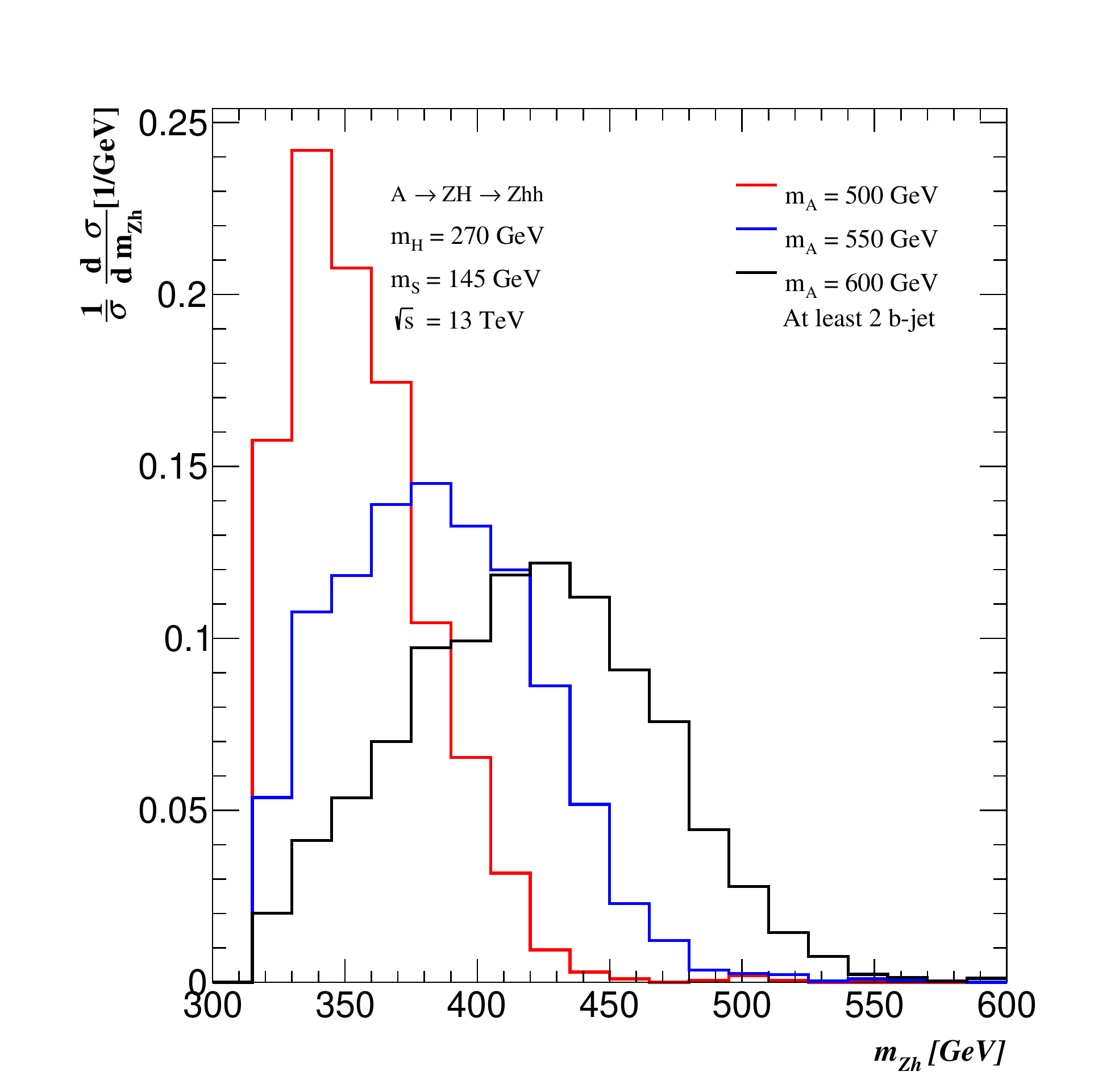}
\includegraphics[width=18pc,height=17pc]{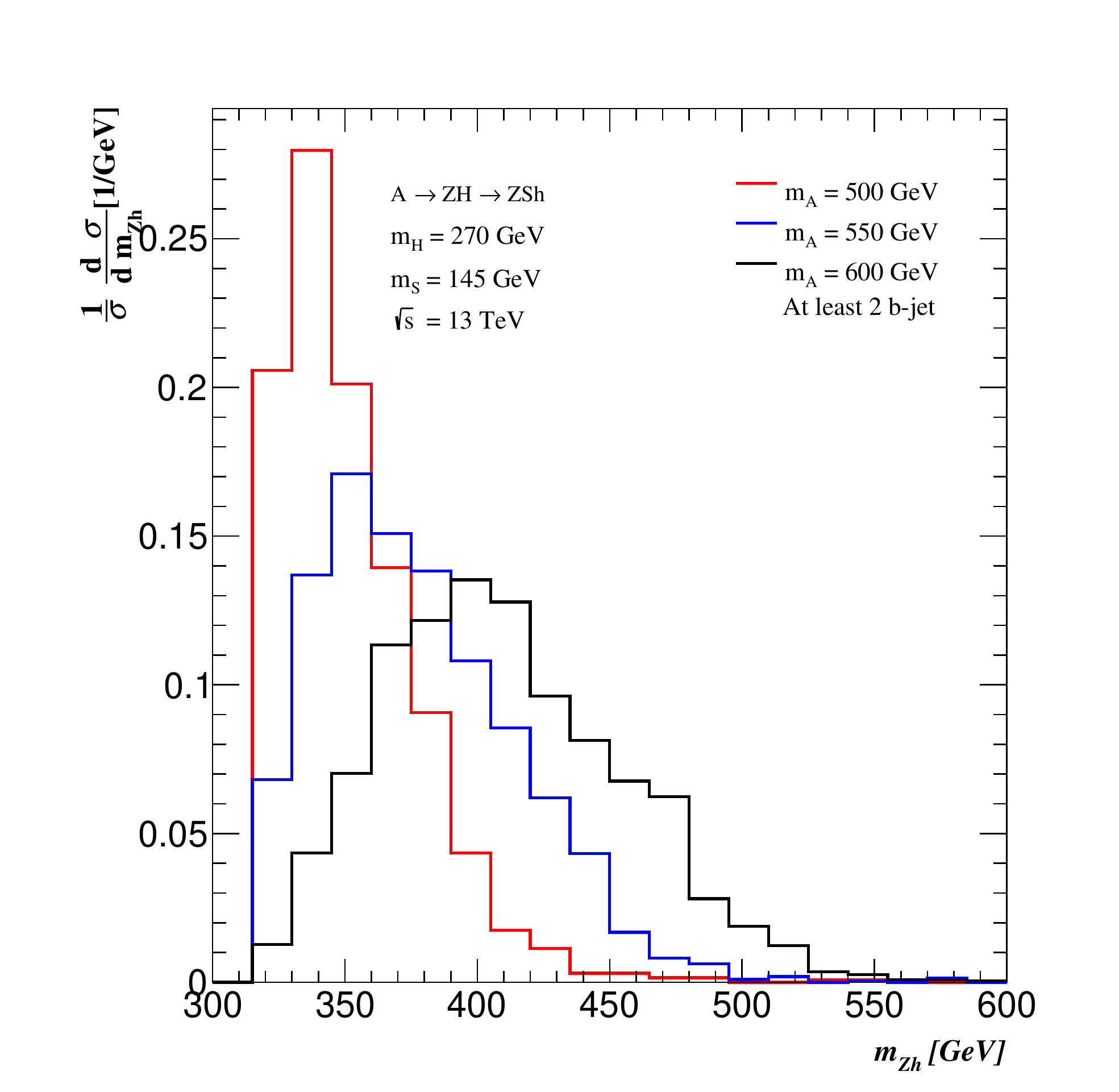}
\caption{Distribution of the invariant mass of the $Zh$ system from the $A\to ZH\to Zhh$ ($left$) and $A\to ZH\to ZSh$ ($right$) decays, where $m_H=270$~GeV and $m_S=145$~GeV. Results are shown for different pseudo-scalar masses after the event selection described in section~\ref{sec:atozh}.}
\label{fig:app}
\end{figure}

Figure~\ref{fig:app} displays the distribution of the invariant mass of the $Zh$ system from the $A\to ZH\to Zhh$ and $A\to ZH\to ZSh$ decays after the application of the event selection detailed in section~\ref{sec:atozh}. The distributions display a cut-off at 320\,GeV described in section~\ref{sec:atozh}. As the $m_A$ becomes larger so that the phase-space available enhancing the invariant mass of the $Zh$ system and its apparent width. 

\ack
The authors would like to thank the DST for the support through the SA-CERN consortium and the NRF for different types of support. The authors would also like to acknowledge the support from the Research Office of the University of the Witwatersrand. E.~Iarilala acknowledges the support from the DAAD program. E.~Shrif acknowledges the support from the Faculty for the Future Schlumberger Foundation. The authors would like to thank Toby Opferkuch for assistance with the model files.

\pagebreak
%\bibliographystyle{spphys}
%{\small\bibliography{references.bib}}
%\section*{References}
%\begin{thebibliography}{9}

\References

%\cite{vonBuddenbrock:2017gvy}
\bibitem{vonBuddenbrock:2017gvy} 

  S.~von Buddenbrock, A.~S.~Cornell, A.~Fadol, M.~Kumar, B.~Mellado and X.~Ruan,
  %``Multi-lepton signatures of additional scalar bosons beyond the Standard Model at the LHC,''
  J.\ Phys.\ G {\bf 45}, no. 11, 115003 (2018)
  %doi:10.1088/1361-6471/aae3d6
  [arXiv:1711.07874 [hep-ph]].
  %%CITATION = doi:10.1088/1361-6471/aae3d6;%%
  %5 citations counted in INSPIRE as of 15 Mar 2019

\bibitem{Englert:1964et}
  F.~Englert and R.~Brout,
  %``Broken Symmetry and the Mass of Gauge Vector Mesons,''
  Phys.\ Rev.\ Lett.\  {\bf 13} (1964) 321.
  
 \bibitem{Higgs:1964pj}
  P.~W.~Higgs,
  %``Broken Symmetries and the Masses of Gauge Bosons,''
  Phys.\ Rev.\ Lett.\  {\bf 13} (1964) 508.

\bibitem{Higgs:1964ia}
  P.~W.~Higgs,
  %``Broken symmetries, massless particles and gauge fields,''
  Phys.\ Lett.\  {\bf 12} (1964) 132.  
  
  \bibitem{Guralnik:1964eu}
  G.~S.~Guralnik, C.~R.~Hagen and T.~W.~B.~Kibble,
  %``Global Conservation Laws and Massless Particles,''
  Phys.\ Rev.\ Lett.\  {\bf 13} (1964) 585.

\bibitem{Aad:2012tfa}
  G.~Aad {\it et al.} [ATLAS Collaboration],
  %``Observation of a new particle in the search for the Standard Model Higgs boson with the ATLAS detector at the LHC,''
  Phys.\ Lett.\ B {\bf 716} (2012) 1.
  
  \bibitem{Chatrchyan:2012xdj}
  S.~Chatrchyan {\it et al.} [CMS Collaboration],
  %``Observation of a new boson at a mass of 125 GeV with the CMS experiment at the LHC,''
  Phys.\ Lett.\ B {\bf 716} (2012) 30.

%\cite{vonBuddenbrock:2015ema}
\bibitem{vonBuddenbrock:2015ema} 
  S.~von Buddenbrock {\it et al.},
  %``The compatibility of LHC Run 1 data with a heavy scalar of mass around 270\,GeV,''
  arXiv:1506.00612 [hep-ph].
  %%CITATION = ARXIV:1506.00612;%%
  %19 citations counted in INSPIRE as of 09 Nov 2017

%\cite{Kumar:2016vut}
\bibitem{Kumar:2016vut} 
  M.~Kumar {\it et al.},
  %``The impact of additional scalar bosons at the LHC,''
  J.\ Phys.\ Conf.\ Ser.\  {\bf 802}, no. 1, 012007 (2017)
  [arXiv:1603.01208 [hep-ph]].
  %%CITATION = doi:10.1088/1742-6596/802/1/012007;%%
  %8 citations counted in INSPIRE as of 09 Nov 2017

 %\cite{vonBuddenbrock:2016rmr}
\bibitem{vonBuddenbrock:2016rmr} 
  S.~von Buddenbrock {\it et al.},
  %``Phenomenological signatures of additional scalar bosons at the LHC,''
  Eur.\ Phys.\ J.\ C {\bf 76}, no. 10, 580 (2016)
  [arXiv:1606.01674 [hep-ph]].
  %%CITATION = doi:10.1140/epjc/s10052-016-4435-8;%%
  %30 citations counted in INSPIRE as of 09 Nov 2017 

%\cite{Fang:2017tmh}
\bibitem{Fang:2017tmh}
  Y.~Fang, M.~Kumar, B.~Mellado, Y.~Zhang and M.~Zhu,
  %``Impact of additional bosons on the exploration of the Higgs boson at the LHC,''
  Int.\ J.\ Mod.\ Phys.\ A {\bf 32} (2017) no.34,  1746010
  [arXiv:1706.06659 [hep-ph]].

%\cite{Mosomane:2017jcg}
\bibitem{Mosomane:2017jcg} 
  C.~Mosomane, M.~Kumar, A.~S.~Cornell and B.~Mellado,
  %``Exploring CP-even scalars of a Two Higgs-doublet model in future $e^-\,p$ colliders,''
  J.\ Phys.\ Conf.\ Ser.\  {\bf 889}, no. 1, 012004 (2017)
  [arXiv:1707.05997 [hep-ph]].
  %%CITATION = doi:10.1088/1742-6596/889/1/012004;%%
  %1 citations counted in INSPIRE as of 20 Aug 2018
  
  %\cite{DelleRose:2018ndz}
\bibitem{DelleRose:2018ndz}
  L.~Delle Rose, O.~Fischer and A.~Hammad,
  %``Prospects for Heavy Scalar Searches at the LHeC,''
  arXiv:1809.04321 [hep-ph].
  %%CITATION = ARXIV:1809.04321;%%
  
%\cite{Das:2018vuk}
\bibitem{Das:2018vuk}
  S.~P.~Das, J.~Hernandez-Sanchez, S.~Moretti and A.~Rosado,
  %``Prospects for discovering a light charged Higgs boson within the NMSSM at the FCC-eh collider,''
  arXiv:1806.08361 [hep-ph].
  %%CITATION = ARXIV:1806.08361;%%

%\cite{Lee:1973iz}
\bibitem{Lee:1973iz} 
  T.~D.~Lee,
  %``A Theory of Spontaneous T Violation,''
  Phys.\ Rev.\ D {\bf 8}, 1226 (1973).
  %doi:10.1103/PhysRevD.8.1226.
  %%CITATION = doi:10.1103/PhysRevD.8.1226;%%
  %1203 citations counted in INSPIRE as of 16 Feb 2019
  
%\cite{Branco:2011iw}
\bibitem{Branco:2011iw} 
  G.~C.~Branco, P.~M.~Ferreira, L.~Lavoura, M.~N.~Rebelo, M.~Sher and J.~P.~Silva,
  %``Theory and phenomenology of two-Higgs-doublet models,''
  Phys.\ Rept.\  {\bf 516}, 1 (2012)
  [arXiv:1106.0034 [hep-ph]].
  %%CITATION = doi:10.1016/j.physrep.2012.02.002;%%
  %1285 citations counted in INSPIRE as of 20 Aug 2018

%\cite{Muhlleitner:2017dkd}
\bibitem{Muhlleitner:2017dkd} 
  M.~Mühlleitner, M.~O.~P.~Sampaio, R.~Santos and J.~Wittbrodt,
  %``Phenomenological Comparison of Models with Extended Higgs Sectors,''
  JHEP {\bf 1708}, 132 (2017)
 % doi:10.1007/JHEP08(2017)132
  [arXiv:1703.07750 [hep-ph]].
  %%CITATION = doi:10.1007/JHEP08(2017)132;%%
  %20 citations counted in INSPIRE as of 14 Mar 2019

%\cite{vonBuddenbrook:2017eqe}
\bibitem{vonBuddenbrook:2017eqe} 
  S.~von Buddenbrook, A.~S.~Cornell, D.~Kar, M.~Kumar, B.~Mellado and R.~G.~Reed,
  %``A heavy scalar of mass 270 GeV and its possible connection to the 750 GeV excess,''
  J.\ Phys.\ Conf.\ Ser.\  {\bf 802}, no. 1, 012001 (2017).
  %%CITATION = doi:10.1088/1742-6596/802/1/012001;%%
  
%\cite{vonBuddenbrock:2017jqp}
\bibitem{vonBuddenbrock:2017jqp} 
  S.~von Buddenbrock, A.~S.~Cornell, M.~Kumar and B.~Mellado,
  %``The Madala hypothesis with Run 1 and 2 data at the LHC,''
  J.\ Phys.\ Conf.\ Ser.\  {\bf 889}, no. 1, 012020 (2017)
  [arXiv:1709.09419 [hep-ph]].
  %%CITATION = doi:10.1088/1742-6596/889/1/012020;%%
  
\bibitem{Mellado_HDAYS2018} 
B.~Mellado, 
Understanding the production of multiple leptons at the LHC, HDAYS2018, Santander, September 10$^{th}$-14$^{th}$ 2018, 
\url{http://hdays.csic.es/HDays18/}.

%\cite{vonBuddenbrock:2019ajh}
\bibitem{vonBuddenbrock:2019ajh} 
  S.~von Buddenbrock, A.~S.~Cornell, Y.~Fang, A.~Fadol Mohammed, M.~Kumar, B.~Mellado and K.~G.~Tomiwa,
  %``The emergence of multi-lepton anomalies at the LHC and their compatibility with new physics at the EW scale,''
  arXiv:1901.05300 [hep-ph].
  %%CITATION = ARXIV:1901.05300;%%
  
\bibitem{Aaboud:2017cxo}
  M.~Aaboud {\it et al.} [ATLAS Collaboration],
  %``Search for heavy resonances decaying into a $W$ or $Z$ boson and a Higgs boson in final states with leptons and $b$-jets in 36 fb$^{-1}$ of $\sqrt s = 13$ TeV $pp$ collisions with the ATLAS detector,''
  JHEP {\bf 1803} (2018) 174
  [arXiv:1712.06518 [hep-ex]].

\bibitem{Khachatryan:2015lba}
  V.~Khachatryan {\it et al.} [CMS Collaboration],
  %``Search for a pseudoscalar boson decaying into a Z boson and the 125 GeV Higgs boson in $?^+?^?b\overline{b}$ final states,''
  Phys.\ Lett.\ B {\bf 748} (2015) 221
  [arXiv:1504.04710 [hep-ex]].
  %%CITATION = doi:10.1016/j.physletb.2015.07.010;%%
  %62 citations counted in INSPIRE as of 23 Nov 2017    
    
%\cite{CMS:2018xvc}
\bibitem{CMS:2018xvc}
  CMS Collaboration [CMS Collaboration],
  %``Search for a heavy pseudoscalar boson decaying to a Z boson and a Higgs boson at sqrt(s)=13 TeV,''
  CMS-PAS-HIG-18-005.
  %%CITATION = CMS-PAS-HIG-18-005;%%
  %6 citations counted in INSPIRE as of 29 Nov 2018    
    
\bibitem{Ferreira:2017bnx}
  P.~M.~Ferreira, S.~Liebler and J.~Wittbrodt,
  %``$pp\to A\to Zh$ and the wrong-sign limit of the two-Higgs-doublet model,''
  Phys.\ Rev.\ D {\bf 97} (2018) no.5,  055008
  [arXiv:1711.00024 [hep-ph]].  

\bibitem{Sirunyan:2017leh}
  A.~M.~Sirunyan {\it et al.} [CMS Collaboration],
  %``Search for top squarks and dark matter particles in opposite-charge dilepton final states at $\sqrt{s}=$ 13 TeV,''
  Phys.\ Rev.\ D {\bf 97} (2018) no.3,  032009
  [arXiv:1711.00752 [hep-ex]].    
        
\bibitem{Sirunyan:2017uzs}
  A.~M.~Sirunyan {\it et al.} [CMS Collaboration],
  %``Measurement of the cross section for top quark pair production in association with a W or Z boson in proton-proton collisions at $ \sqrt{s}=13 $ TeV,''
  JHEP {\bf 1808} (2018) 011
  [arXiv:1711.02547 [hep-ex]].  
  
%\cite{Muhlleitner:2016mzt}
\bibitem{Muhlleitner:2016mzt} 
  M.~Muhlleitner, M.~O.~P.~Sampaio, R.~Santos and J.~Wittbrodt,
  %``The N2HDM under Theoretical and Experimental Scrutiny,''
  JHEP {\bf 1703}, 094 (2017)
  [arXiv:1612.01309 [hep-ph]].
  %%CITATION = doi:10.1007/JHEP03(2017)094;%%
  %7 citations counted in INSPIRE as of 22 Nov 2017

%\cite{Ivanov:2017dad}
\bibitem{Ivanov:2017dad} 
  I.~P.~Ivanov,
  %``Building and testing models with extended Higgs sectors,''
  Prog.\ Part.\ Nucl.\ Phys.\  {\bf 95}, 160 (2017)
  %doi:10.1016/j.ppnp.2017.03.001
  [arXiv:1702.03776 [hep-ph]].
  %%CITATION = doi:10.1016/j.ppnp.2017.03.001;%%
  %58 citations counted in INSPIRE as of 06 Jul 2019

%\cite{Chen:2013jvg}
\bibitem{Chen:2013jvg} 
  C.~Y.~Chen, M.~Freid and M.~Sher,
  %``Next-to-minimal two Higgs doublet model,''
  Phys.\ Rev.\ D {\bf 89}, no. 7, 075009 (2014)
  [arXiv:1312.3949 [hep-ph]].
  %%CITATION = doi:10.1103/PhysRevD.89.075009;%%
  %13 citations counted in INSPIRE as of 22 Nov 2017

%\cite{Drozd:2014yla}
\bibitem{Drozd:2014yla} 
  A.~Drozd, B.~Grzadkowski, J.~F.~Gunion and Y.~Jiang,
  %``Extending two-Higgs-doublet models by a singlet scalar field - the Case for Dark Matter,''
  JHEP {\bf 1411}, 105 (2014)
  [arXiv:1408.2106 [hep-ph]].
  %%CITATION = doi:10.1007/JHEP11(2014)105;%%
  %33 citations counted in INSPIRE as of 22 Nov 2017

%\cite{Haber:1999zh}
\bibitem{Haber:1999zh} 
  H.~E.~Haber and H.~E.~Logan,
  %``Radiative corrections to the Z b anti-b vertex and constraints on extended Higgs sectors,''
  Phys.\ Rev.\ D {\bf 62}, 015011 (2000)
  %doi:10.1103/PhysRevD.62.015011
  [hep-ph/9909335].
  %%CITATION = doi:10.1103/PhysRevD.62.015011;%%
  %168 citations counted in INSPIRE as of 22 Jun 2019

%\cite{Deschamps:2009rh}
\bibitem{Deschamps:2009rh} 
  O.~Deschamps, S.~Descotes-Genon, S.~Monteil, V.~Niess, S.~T'Jampens and V.~Tisserand,
  %``The Two Higgs Doublet of Type II facing flavour physics data,''
  Phys.\ Rev.\ D {\bf 82}, 073012 (2010)
  %doi:10.1103/PhysRevD.82.073012
  [arXiv:0907.5135 [hep-ph]].
  %%CITATION = doi:10.1103/PhysRevD.82.073012;%%
  %154 citations counted in INSPIRE as of 22 Jun 2019

%\cite{Mahmoudi:2009zx}
\bibitem{Mahmoudi:2009zx} 
  F.~Mahmoudi and O.~Stal,
  %``Flavor constraints on the two-Higgs-doublet model with general Yukawa couplings,''
  Phys.\ Rev.\ D {\bf 81}, 035016 (2010)
  %doi:10.1103/PhysRevD.81.035016
  [arXiv:0907.1791 [hep-ph]].
  %%CITATION = doi:10.1103/PhysRevD.81.035016;%%
  %244 citations counted in INSPIRE as of 22 Jun 2019

%\cite{Hermann:2012fc}
\bibitem{Hermann:2012fc} 
  T.~Hermann, M.~Misiak and M.~Steinhauser,
  %``$\bar{B}\to X_s \gamma$ in the Two Higgs Doublet Model up to Next-to-Next-to-Leading Order in QCD,''
  JHEP {\bf 1211}, 036 (2012)
  %doi:10.1007/JHEP11(2012)036
  [arXiv:1208.2788 [hep-ph]].
  %%CITATION = doi:10.1007/JHEP11(2012)036;%%
  %170 citations counted in INSPIRE as of 22 Jun 2019

%\cite{Misiak:2015xwa}
\bibitem{Misiak:2015xwa} 
  M.~Misiak {\it et al.},
  %``Updated NNLO QCD predictions for the weak radiative B-meson decays,''
  Phys.\ Rev.\ Lett.\  {\bf 114}, no. 22, 221801 (2015)
  %doi:10.1103/PhysRevLett.114.221801
  [arXiv:1503.01789 [hep-ph]].
  %%CITATION = doi:10.1103/PhysRevLett.114.221801;%%
  %265 citations counted in INSPIRE as of 22 Jun 2019

%\cite{Coimbra:2013qq}
\bibitem{Coimbra:2013qq} 
  R.~Coimbra, M.~O.~P.~Sampaio and R.~Santos,
  %``ScannerS: Constraining the phase diagram of a complex scalar singlet at the LHC,''
  Eur.\ Phys.\ J.\ C {\bf 73}, 2428 (2013)
  [arXiv:1301.2599 [hep-ph]].
  %%CITATION = doi:10.1140/epjc/s10052-013-2428-4;%%
  %44 citations counted in INSPIRE as of 09 Jul 2018
  
    \bibitem{Mellado_HDAYS2017}
  B.~Mellado, ``The production of additional bosons and the impact on the
Large Hadron Collider",  HDAYS, Santander 18$^{th}$-22$^{th}$ September 2017, 
\url{http://hdays.csic.es/HDays17/}.

%\cite{ATL-PHYS-PUB-2014-006}
\bibitem{ATL-PHYS-PUB-2014-006} 
  G. Aad {\it et al.} [ATLAS Collaboration], ``Update of the prospects for the $H \to Z \gamma$ search at the High-Luminosity LHC'',
  ATL-PHYS-PUB-2014-006,
  \url{http://cds.cern.ch/record/1703276}.  
  
 %\cite{Sjostrand:2014zea}
\bibitem{Sjostrand:2014zea} 
  T.~Sjöstrand {\it et al.},
  %``An Introduction to PYTHIA 8.2,''
  Comput.\ Phys.\ Commun.\  {\bf 191}, 159 (2015)
  %doi:10.1016/j.cpc.2015.01.024
  [arXiv:1410.3012 [hep-ph]].
  %%CITATION = doi:10.1016/j.cpc.2015.01.024;%%
  %1484 citations counted in INSPIRE as of 14 Mar 2019
  
 %\cite{Ovyn:2009tx}
 \bibitem{Ovyn:2009tx} 
 S.~Ovyn, X.~Rouby and V.~Lemaitre,
 %``DELPHES, a framework for fast simulation of a generic collider experiment,''
 [arXiv:0903.2225 [hep-ph]].
 %%CITATION = ARXIV:0903.2225;%%
 %333 citations counted in INSPIRE as of 19 Aug 2018
 
 %\cite{Cacciari:2011ma}
 \bibitem{Cacciari:2011ma} 
 M.~Cacciari, G.~P.~Salam and G.~Soyez,
 %``FastJet User Manual,''
 Eur.\ Phys.\ J.\ C {\bf 72}, 1896 (2012),
 [arXiv:1111.6097 [hep-ph]].
 %%CITATION = doi:10.1140/epjc/s10052-012-1896-2;%%
 %2482 citations counted in INSPIRE as of 19 Aug 2018
 
 %\cite{Cacciari:2008gp}
 \bibitem{Cacciari:2008gp} 
 M.~Cacciari, G.~P.~Salam and G.~Soyez,
 %``The Anti-k(t) jet clustering algorithm,''
 JHEP {\bf 0804}, 063 (2008),
 [arXiv:0802.1189 [hep-ph]].
 %%CITATION = doi:10.1088/1126-6708/2008/04/063;%%
 %5443 citations counted in INSPIRE as of 19 Aug 2018
      
  \bibitem{Aaboud:2017xsd}
  M.~Aaboud {\it et al.} [ATLAS Collaboration],
  %``Evidence for the $ H\to b\overline{b} $ decay with the ATLAS detector,''
  JHEP {\bf 1712} (2017) 024
  [arXiv:1708.03299 [hep-ex]].

    \bibitem{ATLAS:2018nkp}
  The ATLAS collaboration [ATLAS Collaboration],
  %``Observation of $H \to b\bar{b}$ decays and $VH$ production with the ATLAS detector,''
  ATLAS-CONF-2018-036.
  
\bibitem{Sirunyan:2017elk}
  A.~M.~Sirunyan {\it et al.} [CMS Collaboration],
  %``Evidence for the Higgs boson decay to a bottom quark?antiquark pair,''
  Phys.\ Lett.\ B {\bf 780} (2018) 501
  [arXiv:1709.07497 [hep-ex]].  
  
  \bibitem{CMS:2018abb}
  CMS Collaboration [CMS Collaboration],
  %``Observation of Higgs boson decay to bottom quarks,''
  CMS-PAS-HIG-18-016.

\bibitem{Sirunyan:2019wph}
  A.~M.~Sirunyan {\it et al.} [CMS Collaboration],
  %``Search for heavy Higgs bosons decaying to a top quark pair in proton-proton collisions at $\sqrt{s} =$ 13 TeV,''
  arXiv:1908.01115 [hep-ex].

%\cite{CMS:2019wml}
\bibitem{CMS:2019wml} 
  CMS Collaboration [CMS Collaboration],
  %``Search for 2HDM neutral Higgs bosons through the $\mathrm{H} \to \mathrm{Z}\mathrm{A} \to \ell^{+}\ell^{-}\mathrm{b}\overline{\mathrm{b}}$ process in proton-proton collisions at $\sqrt{s} = 13~\mathrm{TeV}$,''
  CMS-PAS-HIG-18-012.
  %%CITATION = CMS-PAS-HIG-18-012;%%

%\cite{Aaboud:2018eoy}
\bibitem{Aaboud:2018eoy} 
  M.~Aaboud {\it et al.} [ATLAS Collaboration],
  %``Search for a heavy Higgs boson decaying into a $Z$ boson and another heavy Higgs boson in the $\ell\ell bb$ final state in $pp$ collisions at $\sqrt{s}=13$ TeV with the ATLAS detector,''
  Phys.\ Lett.\ B {\bf 783}, 392 (2018)
  %doi:10.1016/j.physletb.2018.07.006
  [arXiv:1804.01126 [hep-ex]].
  %%CITATION = doi:10.1016/j.physletb.2018.07.006;%%
  %23 citations counted in INSPIRE as of 04 Jul 2019

\endrefs        
% \end{thebibliography}
\end{document}